\documentclass[12pt]{article}

\title{Proof of the Ergodic Theorem and the $H$-Theorem in Quantum Mechanics\footnote{Translation of: Beweis des Ergodensatzes und des $H$-Theorems in der neuen Mechanik. 
      \textit{Zeitschrift f\"ur Physik} \textbf{57}: 30--70 (1929). 
Translated in 2009. Additions in the text by the translator are put in square brackets. Footnotes are also footnotes in the original unless otherwise marked. Footnotes in the original containing only citations have been moved to the main text. In the original, equations and references are not numbered. The notation agrees essentially with the original, with the following exceptions: $h/2\pi$ has been replaced with $\hbar$; the notation $[a,b]$ for intervals has been introduced to simplify some sentences. In a few cases, misprints and other mistakes in formulas have been identified by the translator, corrected in the text, and mentioned in a footnote. The translator is grateful to Wolf Beiglb\"ock for suggesting improvements and librarian Mei Ling Lo of Rutgers University for help with the bibliography.}}
\author{Author: J. von Neumann (Berlin)\\Translator: Roderich Tumulka\footnote{Department of Mathematics, Rutgers University, 110 Frelinghuysen Road, Piscataway, NJ 08854-8019, USA. E-mail: tumulka@math.rutgers.edu}}

\date{received May 10, 1929}
\usepackage{amssymb,amsmath,url}

\DeclareMathOperator{\tr}{tr}
\newcommand{\1}{\boldsymbol{\mathsf{1}}}
\newcommand{\DDelta}{\boldsymbol{\mathsf{\Delta}}}
\newcommand{\AAA}{\boldsymbol{\mathsf{A}}}
\newcommand{\BBB}{\boldsymbol{\mathsf{B}}}
\newcommand{\EEE}{\boldsymbol{\mathsf{E}}}
\newcommand{\HHH}{\boldsymbol{\mathsf{H}}}
\newcommand{\QQQ}{\boldsymbol{\mathsf{Q}}}
\newcommand{\PPP}{\boldsymbol{\mathsf{P}}}
\newcommand{\UUU}{\boldsymbol{\mathsf{U}}}
\newcommand{\M}{\mathfrak{M}}
\newcommand{\Mb}{\boldsymbol{\mathrm{M}}}
\newcommand{\Nb}{\boldsymbol{\mathrm{N}}}
\newcommand{\be}{\begin{equation}}
\newcommand{\ee}{\end{equation}}

\begin{document}

\maketitle

\begin{abstract}
It is shown how to resolve the apparent contradiction between the macroscopic approach of phase space and the validity of the uncertainty relations. The main notions of statistical mechanics are re-interpreted in a quantum-mechanical way, the ergodic theorem and the $H$-theorem are formulated and proven (without ``assumptions of disorder''), followed by a discussion of the physical meaning of the mathematical conditions characterizing their domain of validity.
\end{abstract}
\setcounter{section}{-1}
\section{Introduction}

\subsection{}\label{sec:1}

The object of the present paper is the clarification of the relations between the macroscopic and the microscopic point of view of complex systems; that is, the discussion of the question why the known thermodynamic methods of statistical mechanics make it possible to make statements about incompletely (viz., only macroscopically) known systems that are correct most of the time. In particular, first, how the peculiar, seemingly irreversible behavior of entropy arises, and second, why the statistical properties of the (fictitious) micro-canonical ensemble can be attributed to the incompletely known (real) system.\footnote{We are thinking of closed and isolated systems. For a system in contact with a large heat reservoir it is known that the so-called canonical ensemble is appropriate. However, this case can easily be reduced, with the methods of statistical mechanics, to the former, by including the heat reservoir into the system.} And these questions shall be attacked with the means of quantum mechanics.

In classical mechanics, it is known that these questions have led to the development of two elaborate theoretical systems: the statistical mechanics of Boltzmann and that of Gibbs. The former could not provide a final and satisfactory solution because it had to make essential use of so-called assumptions of disorder---and exactly to fathom the nature of this ``disorder'' is the real problem.\footnote{For a critical discussion of this matter (also concerning our subsequent remarks) see \cite{Ehr,Ehr07}.} The latter would basically be adequate for this program; however, it leads to a mathematical problem---the so-called quasi-ergodic problem---that has been and still is absolutely insurmountable. Only if the corresponding mathematical conjecture is valid, the Gibbsian theory succeeds.

In general questions of principle, however, the new quantum mechanics differs from the classical mechanics by being remarkably simple;\footnote{For many special problems it is, of course, the other way around.} it is due to this circumstance that in quantum mechanics, if we follow the Gibbsian path, we can reach the goal with relatively simple mathematical means. That is, it will be possible in what follows to prove the ergodic theorem and the $H$-theorem (which are the two questions mentioned above) without the need to recur to any assumption of disorder. But before speaking of them in more detail, we need to say more about the notion of the macroscopic in quantum mechanics.

\subsection{}\label{sec:2}

The main difficulty with re-constructing the Gibbsian theory in quantum mechanics is that the tool of ``phase space''---i.e., for a system of $f$ degrees of freedom, the $2f$-dimensional space described by the $f$ coordinates $q_1,\ldots,q_f$ and the $f$ momenta $p_1,\ldots,p_f$---cannot be dispensed with: all of the important notions (energy surface, phase cells, micro-canonical and canonical ensembles, etc.) are based on it. But the phase space cannot be formed in quantum mechanics, since a coordinate $q_k$ and the corresponding momentum $p_k$ are never simultaneously measurable; instead, their probable errors (spreads) $\Delta q_k$ and $\Delta p_k$ are always related according to the uncertainty relation $\Delta q_k \, \Delta p_k \geq  \hbar/2$.\footnote{\label{fn:++31}See \cite{Hei27} and \cite{Bohr28}. Concerning the limit $\hbar/2$ see, e.g., \cite[p.~272]{Weyl28}.} Moreover, it is impossible to specify, for a state of the system, two intervals $I,J$ so that, with certainty, $q_k$ lies in $I$ and $p_k$ in $J$ (even if the product of their lengths is much bigger than $\hbar/2$)\footnote{\label{fn:+++31}That is, if the wave function $\varphi(q_1,\ldots,q_f)$ vanishes for all values of $q_k$ outside a finite interval $I$ then, expanding 
\[
\varphi(q_1,\ldots,q_f) = \int_{-\infty}^\infty \cdots \int_{-\infty}^\infty c(p_1,\ldots, p_f) e^{\frac{i}{\hbar} (p_1 q_1 + \ldots + p_f q_f)} dp_1 \cdots dp_f,
\]
the Fourier coefficients $c(p_1,\ldots,p_f)$ must again and again become $\neq 0$ for arbitrarily large $p_k$.}---thus, not only the continuous phase space but also a discrete partition thereof into cells is meaningless! Still, it is obviously factually correct that in macroscopic measurements the coordinates and momenta are measured simultaneously---indeed, the idea is that that becomes possible through the inaccuracy of the macroscopic measurement, which is so great that we need not fear a conflict with the uncertainty relations. How are these two statements, contradicting each other, to be reconciled?

We believe that the following interpretation is the correct one: In a macroscopic measurement of coordinate and momentum (or two other quantities that cannot be measured simultaneously according to quantum mechanics), really two physical quantities are measured simultaneously and exactly, which however are not exactly coordinate and momentum. They are, for example, the orientations of two pointers or the locations of two spots on photographic plates\footnote{For example, one may think of the coordinate and momentum of a particle in the sense of the citations of Footnote~\ref{fn:++31} as measured in the following way: On the one hand (coordinate), let the particle be illuminated by a bundle of light focussed on it approximate position, on the other hand (momentum) by a quite monochromatic and plane wave bundle of light, with the reflected light photographed after passing a prism in order to determine the wave length. Of course, the inaccuracies must satisfy the uncertainty relation. In this way one obtains, on two photographic plates, two spots determining coordinate and momentum with said inaccuracy.}---and nothing keeps us from measuring these simultaneously and with arbitrary accuracy, only their relation to the really interesting physical quantities ($q_k$ and $p_k$) is somewhat loose, namely the uncertainty of this coupling required by the laws of nature corresponds to the uncertainty relation (cf.~Footnote~\ref{fn:++31}).

Formulated mathematically, quantum mechanics attributes to the quantities $q_k$ and $p_k$ the well-known operators $\QQQ_k=q_k\cdots$ and $\PPP_k=\frac{\hbar}{i} \frac{\partial}{\partial q_k} \cdots$ \cite{Schr26}, whose lack of commutability ($\QQQ_k\PPP_k \neq \PPP_k \QQQ_k$, the difference is, as is well known, $\frac{\hbar}{i} 1$) corresponds to the lack of simultaneous measurability of these quantities \cite{Dir27,Hei27}. We now assume that two other, commuting, operators $\QQQ'_k, \PPP'_k$ exist whose difference from $\QQQ_k$ (respectively, $\PPP_k$) is so small that its size is characterized by numbers $\Delta Q_k$ and $\Delta P_k$ whose product does not significantly exceed the value $\hbar/2$ required by the uncertainty relation. (Of course, it cannot be less than that because of $\QQQ_k\PPP_k - \PPP_k \QQQ_k = \frac{\hbar}{2i} 1$, $\QQQ'_k\PPP'_k - \PPP'_k\QQQ'_k=0$!) A somewhat different formulation that achieves (as one easily sees) the same arises from the following consideration: The commuting operators $\QQQ'_k,\PPP'_k$ must possess a complete orthogonal system of common eigenfunctions,\footnote{\label{fn:33+}For the sake of simplicity, we assume that the actually measured quantities $Q'_k,P'_k$ have pure point spectra, which should be the case if the available volume is finite. The existence of a system of common eigenfunctions can be proved in the same way as for usual (finite dimensional) matrices \cite{Fro77,HT}.} denoted $\varphi_1,\varphi_2,\ldots$. Thereof we have to require that in every state $\varphi_n$ the spreads of $\QQQ_k$ and $\PPP_k$ are less than $\Delta Q_k$ and $\Delta P_k$ (where $\Delta Q_k \, \Delta P_k \sim \hbar/2$). Then a simultaneous measurement of $\QQQ'_k$ and $\PPP'_k$, which must lead to a state $\varphi_n$, does indeed provide simultaneous information about $\QQQ_k$ and $\PPP_k$. By the way, it suffices to select the orthogonal system $\varphi_1,\varphi_2,\ldots$ as described above, then $\QQQ'_k$ and $\PPP'_k$ can then easily be chosen---after all, it suffices to specify their respective eigenvalues in the states $\varphi_n$ ($n=1,2,\ldots$), which it is advantageous to take to be the expectation values of $\QQQ_k$ and $\PPP_k$ in the state $\varphi_n$.\footnote{That is, $\int_{-\infty}^\infty q_k \, |\varphi_n(q_1\ldots q_f)|^2 \, dq_1 \cdots dq_f$ and $\frac{\hbar}{i} \int_{-\infty}^\infty \varphi'_{q_k}(q_1\ldots q_f) \, \varphi^*(q_1 \ldots q_f)\, dq_1 \cdots dq_f$.}

This plausible assumption can be confirmed mathematically: For any two positive numbers $\varepsilon,\eta$ with $\varepsilon\eta = C\hbar/2$ (where $C$ is a constant, see Footnote~\ref{fn:+++33}), there is a complete orthogonal system $\varphi_1,\varphi_2,\ldots$ such that in every state $\varphi_n$ the spreads of $\QQQ_k$ and $\PPP_k$ are smaller than $\varepsilon$ (respectively, $\eta)$.\footnote{\label{fn:+++33}One sees that $C\approx 1$ would be the ideal estimate (which exploits all possibilities left by the uncertainty relation). The author succeeded only in computing $C<3.6$ [Note of the translator: 3 years later in his book, von Neumann repeated this claim with $C\sim 60$, so maybe the bound $C<3.6$ was incorrectly calculated; see also Section 2.2 of the commentary], but since the value of $\hbar/2$ in macroscopic (centimeter-gram-second) units is approximately $10^{-28}$, the difference does not really matter.} To specify the $\varphi_n$ and to prove their properties requires somewhat cumbersome calculations,\footnote{One should use the wave packets used by Heisenberg \cite{Hei27}, $\exp(-\frac{1}{4\Theta^2}q^2+(\frac{\alpha}{2\Theta^2}+\frac{i}{\hbar}b)q)$---where we write $q$ for $q_k$ and ignore the other $q_1,\ldots, q_f$, so that $\QQQ=q\cdots$ and $\PPP=\frac{\hbar}{i} \frac{\partial}{\partial q} \cdots$ have the means $a$ respectively $b$ and the spread squares $\Theta^2$ respectively $(\frac{\hbar}{2\Theta^2})^2$---with $a=\sqrt{4\pi/C} \varepsilon i$, $b=\sqrt{4\pi/C}\eta j=\sqrt{C\pi}(\hbar/\varepsilon)j$, $\Theta=\varepsilon/\sqrt{C}$, where $i,j = 0, \pm 1, \pm 2, \ldots$. The functions thus defined should be written in arbitrary order as a sequence and then orthogonalized according to the procedure of E.~Schmidt \cite{Schm}. This yields the desired $\varphi_1,\varphi_2,\ldots$.} which we do not reproduce here since the important aspects should be sufficiently clear from the above description.

Thus, we make the assumption about the nature of macroscopic measurements that simultaneously measurable quantities (with pairwise commuting operators) are being measured, which are coupled to the primitive and not simultaneously measurable physical quantities (coordinates, momenta, etc.) just so accurately as allowed by the uncertainty relations. How to carry this out in detail will be shown in the course of this paper.

\subsection{}\label{sec:3}

About the formalism of quantum mechanics in general we say the following. The states of a system are known to be characterized by the so-called wave functions, complex functions $\varphi=\varphi(q_1,\ldots,q_f)$ defined on the ``configuration space'', the $f$-dimensional space described by the $f$ coordinates $q_1,\ldots,q_f$. The physical quantities are characterized by the Hermitian operators $\AAA,\BBB,\ldots$.\footnote{In the following, the terminology and notation follows that of \cite{vN27b}. Everything needed for the present purposes will be collected presently.} The most important operations with wave functions are: the ``inner product''
\be
(\varphi,\psi) = \int \cdots \int \varphi(q_1,\ldots,q_f) \, \psi(q_1,\ldots,q_f)^* \, dq_1\cdots dq_f
\ee
(where $^*$ denotes the complex conjugate) and the ``norm''\footnote{The calculus with these is outlined, e.g., in \cite{vN27a}.}
\be
\|\varphi \| = \sqrt{(\varphi,\varphi)} 
=\sqrt{\int\cdots\int |\varphi(q_1,\ldots,q_f)|^2\, dq_1 \cdots dq_f}\,.
\ee
The simplest description of a state by means of a wave function $\varphi$ is obtained in this way: the expectation value of the quantity $\AAA$ in the state $\varphi$ is equal to $(\AAA\varphi,\varphi)$. The specification of all expectation values provides, as it includes the expectation values of all powers (i.e., the so-called higher moments of a probability distribution), knowledge of the entire probability distribution of every quantity---and thus a complete statistical characterization of the system \cite{Dir27,vN27b}.

We also need the statistics of quantities in the system in case that, instead of a single states $\varphi$, we encounter a mixture of several states $\varphi_1,\varphi_2,\ldots$ with respective probabilities $w_1,w_2,\ldots$. Then the expectation value of $\AAA$ is, obviously, equal to $\sum_n w_n (\AAA\varphi_n,\varphi_n)$, which is advantageously written in a different way. Let us describe, in any complete orthogonal system, $\AAA$ by a matrix $a_{\mu\nu}$ and each $\varphi_n$ by a vector $x^n_\mu$ ($\mu,\nu=1,2,\ldots)$ \cite{vN27a}. Then
\be
\sum_n w_n (\AAA\varphi_n,\varphi_n) = \sum_n w_n \sum_{\mu,\nu} a_{\mu\nu}\, {x^n_\mu}^* \, x^n_\nu = \sum_{\mu,\nu} a_{\mu\nu} \Bigl[\sum_n w_n \, x^n_\nu {x^n_\mu}^* \Bigr]\,,
\ee
so that, if $\UUU$ is the operator with matrix $\sum_n w_n \, x^n_\nu {x^n_\mu}^*$, this is the trace of $\AAA\UUU$.\footnote{\label{fn:+++35}See \cite{vN27b,Dir28}. The trace is the sum of the diagonal elements of the matrix; since it is a unitary invariant, one can talk of the trace of an operator, without specifying a complete orthogonal system.} Thus the statistical behavior of the above mixture of several states is characterized by the operator $\UUU$, on the basis of the rule: the expectation value of $\AAA$ is equal to $\tr(\AAA\UUU)$. We call $\UUU$ the statistical operator of the mixture; one sees that $\UUU$ suffices for describing the mixture, and it is unnecessary to specify the individual states from which it was composed.

By the way, it is convenient to introduce a symbol $\PPP_\varphi$ for the operator with the matrix $x_\mu {x_\nu}^*$ (where $x_\mu$ is the vector of the wave function $\varphi$). It is easy to verify the equivalent definition $\PPP_\varphi f = (f,\varphi) \, \varphi$ (where $f$ is any other wave function). Then, $\UUU=\sum_n w_n \PPP_{\varphi_n}$; in particular, $\PPP_\varphi$ is the statistical operator of the pure state $\varphi$.

\subsection{}\label{sec:4}

Now we can approach the (quantum mechanical) formulation of the ergodic theorem. We start by discussing two approaches that do not solve the real problem but will, we believe, help make the situation clearer and more transparent.

The classical formulation of the ergodic theorem (more precisely, the quasi-ergodic theorem) asserts the following: A system's point in phase space will, in the course of its motion (determined by the differential equations of mechanics), come arbitrarily close to every point of its energy surface---indeed, the time it spends in any region of the latter in the long time average is proportional to the measure of that region.\footnote{\label{fn:+36}As is well known, the measure to be considered is not the $(2f-1)$-dimensional surface area of the piece of energy surface but rather the [infinitesimal] $2f$-dimensional volume of a strip between neighboring energy surfaces, i.e., the integral of the reciprocal [magnitude of the] gradient of the energy over the region mentioned.---The essential (and often ignored) difference between the two halves of the above formulation of the quasi-ergodic theorem was emphasized by P.\ and T.~Ehrenfest \cite{Ehr,Ehr07}: the second half is indispensable for the foundation of the statistical mechanics of Gibbs.} Thus, in a given state the statistical properties of the time ensemble (corresponding to averaging every quantity over all times) are identical to those of its micro-canonical ensemble. The latter is the mixture of all points of the energy surface, with region of equal measure (as in Footnote~\ref{fn:+36}) given equal weight.

Now in quantum mechanics let $\HHH$ be the energy operator, $\varphi_1, \varphi_2,\ldots$ its eigenfunctions,\footnote{More precisely: a complete orthogonal system formed of eigenfunctions, i.e., a coordinate system in which $\HHH$ is diagonal. (We assume that there is no continuous spectrum.)} $W_1,W_2,\ldots$ the respective eigenvalues. A state
\be
\psi = \sum_n a_n \varphi_n
\ee
evolves with time $t$ ([be it] $>0$, $=0$, or $<0$) according to the time-dependent Schr\"odinger equation to
\be\label{SchrEvol}
\psi_t = \sum_n a_n e^{iW_nt/\hbar} \varphi_n= \sum_n a_n(t) \varphi_n\,.
\ee
We first need to scrutinize the concept of energy surface. The $|a_n(t)|^2 = |a_n|^2$ remain constant in the course of time, not only the energy expectation value $(\HHH\psi_t,\psi_t) = \sum_n |a_n(t)|^2 W_n$. Since the $|a_n(t)|^2$ characterize the entire statistics of energy,\footnote{For example, because they determine, according to $(\HHH^k\psi_t,\psi_t) = \sum_n |a_n(t)|^2 W_n^k$, the expectation values of all powers of energy, i.e., all moments of the energy statistics.} we can say: The law of energy conservation in classical mechanics, when transferred to quantum mechanics, asserts not merely the conservation of the mean energy, but rather the conservation of the whole probability distribution of the energy. If we defined a quantum mechanical ``energy surface'' in the immediate way by
\be
\sum_n |a_n|^2 W_n = \mathrm{const.}
\ee
then the ergodic theorem would be far from valid---after all, there exist infinitely many constants of motion $|a_1|^2, |a_2|^2,\ldots$. Instead, the ``energy surface'' should be defined as
\be
|a_1|^2 = \mathrm{const.}_1\,,\quad
|a_2|^2 = \mathrm{const.}_2\,,\quad\ldots.
\ee
We thus arrive at the question: Let
\be
a_n = r_n e^{i\alpha_n} \quad
(r_n \geq 0, \: 0\leq \alpha_n<2\pi),
\ee
so that the energy surface consists of those
\be
\psi' = \sum_n a_n'\varphi_n \quad \text{with} \quad
a_n' = r_n e^{i\alpha_n'} \quad
(0\leq \alpha_n'< 2\pi),
\ee
do the
\be
a_n(t) = r_n e^{i(W_nt/\hbar+\alpha_n)}
\ee
come arbitrarily close to all $a_n'$, i.e., do the $W_nt/\hbar+\alpha_n$ come arbitrarily close to the $\alpha_n'$ (modulo $2\pi$, of course, and for all $n=1,2,\ldots$)? And, how long are the relative sojourn times in given intervals of $\alpha_n'$? Put differently: Will $W_nt/\hbar$ come arbitrarily close, for suitable $t$ and modulo $2\pi$, to any given collection $\alpha_n'-\alpha_n$ (for all $n=1,2,\ldots$), and what are the relative sojourn times? According to theorems of Kronecker, for the former behavior the linear independence of $W_n/\hbar$ over the integers is necessary and sufficient, i.e., the condition that no relation of the form
\be\label{linind}
x_1 \frac{W_1}{\hbar} + \ldots + x_n \frac{W_n}{\hbar}=0
\ee
($n$ arbitrarily large but finite; $x_1,\ldots,x_n$ integers) obtains, except when $x_1=\ldots =x_n=0$ \cite{Kro1884a,Kro1884b}. From further theorems of Weyl it follows that in this case also the sojourn times are correct, i.e., proportional to the product of the lengths of the intervals \cite{Wey15}. So, in this formulation the [hypothesis of the] ergodic theorem amounts to the absence of resonances between the terms $W_n/\hbar$ of the system.\footnote{It may seem strange that the condition involves the $W_n/\hbar$ and not the $(W_m-W_n)/\hbar$, but this is due to an imprecision in our consideration. A constant factor (of modulus 1) in the wave function is meaningless (e.g., it drops out of the statistical operator $\PPP_\psi$), and thus we should have required, what we asked of the phases $W_nt/\hbar + \alpha_n$, only of the phase differences, for example $(W_n-W_1)t/\hbar + (\alpha_n-\alpha_1)$ for $n=2,3,\ldots$. This leads again to condition \eqref{linind} above, but now for the eigen frequencies $(W_n-W_1)/\hbar$, $n=2,3,\ldots$.}

However, we have actually asked too much, as the true essence of the ergodic theorem that is essential to all applications is, as already mentioned, the agreement between the time ensemble and micro-canonical ensemble---and not the question what the system's trajectory on the energy surface is. As we know from Section 0.3, to this end only agreement between the statistical operators of these two ensembles is needed (while, beyond that, their ``true'' composition from wave functions is undiscoverable).

Now $\psi_t$ has the statistical operator $\PPP_{\psi_t}$, and we need to average this, on the one hand, over all $t$ while keeping the $\alpha_n$ fixed (time ensemble), and, on the other hand, for $t=0$ over all $\alpha_n$ (micro-canonical ensemble, where we now write $\alpha_n$ instead of $\alpha_n'$). We want to write $\PPP_{\psi_t}$ as a matrix in the coordinate system $\varphi_1,\varphi_2,\ldots$; since
\be
\psi_t = \sum_n r_n \, e^{i(W_nt/\hbar+\alpha_n)} \varphi_n\,,
\ee
the $m,n$ component of $\PPP_{\psi_t}$ equals
\be
r_m r_n e^{i((W_m-W_n)t/\hbar+ (\alpha_m-\alpha_n))}\,.
\ee
Averaging this over all $\alpha_\ell$, we obtain 0 for $m\neq n$ and $r_m^2$ for $m=n$. To ensure that averaging over $t$ yields the same result, we must have that $(W_m-W_n)/\hbar\neq 0$ for $m\neq n$, i.e., $W_m \neq W_n$. That is, there must not be degeneracies (a much weaker condition than the previous one [i.e., rational-linear independence]!).

At this point we might think we have proved the ergodic theorem to a satisfactory extent. However, we cannot be satisfied with this result since it does not mention the role of the macroscopic. Indeed, we have dealt with a completely and exactly known system, for which, for example, the energy surface was described by the exact specification of all $|a_n|^2$. Thus, in order to treat the incompletely known systems of statistical mechanics, we need to further modify the question we are asking.\footnote{Another hint showing that the theorem just proved cannot be the right ergodic theorem is that its premise (non-degenerate energy) is too weak: it is still satisfied for a known counterexample to the classical ergodic theorem! Cf.\ Section~\ref{sec:III3}.}

\subsection{}\label{sec:5}

This modification must consist primarily in re-interpreting the concept of energy surface in a macroscopic way, i.e., to extend the micro-canonical ensemble to a collection of all those states whose energy statistics cannot macroscopically be distinguished from that of the given state. Under such circumstances, also the agreement between time and microscopic [i.e., micro-canonical] average should only be required for macroscopic quantities. This weakening comes together with an essential strengthening that is made possible only by using the macroscopic perspective. Namely, we will show that for every state of the system the value of each (macroscopically measurable) quantity not only has time mean equal to the micro-canonical mean, but furthermore has small spread, i.e., the times at which the value deviates considerably from the mean are very infrequent.

It is useful to compare this with the corresponding considerations of the classical theory. There, the above-mentioned theorem, which amounts to a justification of the statistical-mechanical methods, gets decomposed into two steps as follows: First it needs to be shown that for every quantity the time statistics coincides with the micro-canonical one; then that for so-called macroscopic quantities the micro-canonical statistics has small spread. The first claim is just the presently unprovable classical quasi-ergodic theorem, the second, in contrast, can easily be proved by means of combinatorial considerations of counting (see, in particular, \cite{Ehr,Ehr07}). However, what we want to call the ergodic theorem is the above implication of both claims together.

A more precise discussion will be provided in the course of this paper; here we just want to emphasize two points: First, our formulation of the ergodic theorem will require that the temporal behavior sketched above actually occurs for every initial state of the system (every $\psi$) \emph{without exceptions} (classically, one would admit exceptions in lower-dimensional parts of the energy surface). Second, we emphasize that the true state (about which we do calculations) is a wave function, i.e., something microscopic---to introduce a macroscopic description of the state would mean to introduce disorder assumptions, which is what we definitely want to avoid. Likewise, the energy operator occurring in the time-dependent Schr\"odinger equation
\be
\frac{\partial}{\partial t}\psi_t = \frac{i}{\hbar} \HHH \psi_t
\ee
(whose solution is \eqref{SchrEvol}) must be represented in its exact (microscopic) form. (Of course, this is different from what happens in the definition of the energy surface, as we will discuss later.) We will now elucidate the conditions that will turn out necessary for the validity of the ergodic theorem.

\subsection{}\label{sec:6}

These conditions come in two groups, first those concerning the (microscopic) energy operator $\HHH$, second those concerning the partition of the (macroscopic) energy surface into phase cells, and the size of the latter. (What is meant quantum-mechanically by energy surface, phase cells, and other objects in phase space, will be defined precisely; at this point it suffice to operate with these terms in the way that was common in pre-quantum-mechanical theory. In particular, by phase cells we mean the partition of phase space that can be carried out by means of macroscopic measurements.)

Concerning the energy, we will find that the term differences (i.e., eigen frequencies) must  be distinct, and likewise the terms themselves (non-degenerate!)---i.e., if $W_1,W_2,\ldots$ are the energy values then all $W_m-W_n$ (with $m\neq n$) are distinct and likewise all $W_n$. (Though we might even admit infrequent exceptions!) As one can see, this condition lies, with respect to its strength, between the two conditions found in Section~\ref{sec:4} [i.e., it is weaker than rational-linear independence and stronger than absence of degeneracies]. We will convince ourselves in Section~\ref{sec:III3} that it is a reasonable condition, in particular one violated by the classical counterexamples to the ergodic theorem (ideal gas without collisions, radiation in a cavity without absorption) and re-instated as valid by the known (but only heuristically confirmed) counteractive measures (introduction of collisions, absorption and emission).

About the size of the phase cells we find the following: \emph{the number of states (quantum orbits) in each phase cell has to be not only very large, but also on average quite large compared to the number of phase cells.} We postpone a more detailed interpretation of this condition until later and mention here only the following: When we take the limit $\hbar\to 0$ (i.e., let quantum mechanics tend to classical mechanics) while not changing the macroscopic measuring technique, then the former number grows unboundedly while the latter is constant---thus, our condition is satisfied better and better. Its validity is thus guaranteed at least if the macroscopic measuring technique is much too coarse to reach quantum effects (so that $\hbar$ is practically 0).

It remains to formulate the $H$-theorem, which we will prove, too. We can attribute in an obvious way an entropy to every state $\psi$, and likewise to its micro-canonical ensemble;\footnote{Cf.\ the end of Section~\ref{sec:I3}, where we will also say more about the relation between this entropy to that defined by the author in \cite{vN27c}.} we can then study the temporal variation of the former and compare it to the latter (which is, as one can easily show, always greater than or equal to the former). As in classical mechanics, also here a monotonic increase of entropy is out of the question, and so is a predominantly positive sign of its [time] derivative (or difference quotient): the time reversal objection as well as the recurrence objection are valid in quantum mechanics as well as in classical mechanics. Following the discussion of P.\ and T.~Ehrenfest of this issue \cite{Ehr,Ehr07}, we see instead the following as the essential statement of the $H$-theorem: the time average of the entropy of $\psi_t$ differs only little from the entropy of the micro-canonical ensemble---and since the latter is an upper bound of the former, we have that the entropy of $\psi_t$ will rarely be much less than the micro-canonical entropy.

We will see that the $H$-theorem holds under the same hypotheses as the ergodic theorem.

To sum up, in quantum mechanics one can prove the ergodic theorem and the $H$-theorem in full rigor and without disorder assumptions; thus, the applicability of the statistical-mechanical methods to thermodynamics is guaranteed without relying on any further hypotheses.\footnote{Cf.\ Schr\"odinger \cite{Schr28}, particularly the last section. Our results allow us to carry out his reasoning in a compelling way without his ``statistical assumption'' (i.e., disorder assumption), and thus to reduce it in full rigor to the ordinary statistical interpretation of quantum mechanics. This also answers Schr\"odinger's question whether quantum mechanics also suffers from an ``ergodic difficulty.''} Of course, this is compatible with the fact that also the time-dependent Schr\"odinger equation, on which quantum mechanics is grounded, has reversibility and recurrence properties just like the differential equations of classical mechanics \cite{Schr28}, and therefore cannot alone explain irreversible phenomena.\footnote{However, quantum mechanics does know an irreversible elementary process: the measurement. It is irreversible (see \cite{vN27c}, where the definition of this process is given in footnote 21 on page 283), but whether it is relevant to the irreversibility of reality we leave open. In this work, we do not deal with measurement.}

\subsection{}\label{sec:7}

We would like to sketch the relation between this work and other quantum-mechanical investigations on questions of statistical mechanics and thermodynamics. The papers of Schr\"odinger \cite{Schr28}, as well as of L.~Nordheim \cite{Nor28}, and W.~Pauli \cite{Pau28} describe the macroscopic situation by means of disorder assumptions, and therefore lie in a different alley of research. An earlier work of the author is based entirely on the microscopic perspective and has the converse goal: To determine the entropy value from assuming the validity of the phenomenological second law of thermodynamics.

The author would like to express his deepest gratitude towards Mr.\ E.~Wigner for numerous discussions in which the questions of this article have arisen.

\section{Quantum-Mechanical Formulation of the Concepts of the Gibbsian Statistical Mechanics}\label{sec:I}

\subsection{}\label{sec:I1}

As we have said and justified in the introduction, we take for granted that all macroscopic observations that are possible at all are possible simultaneously. Thus, their operators all commute with each other, and so there is a complete orthogonal system $\omega_1,\omega_2,\ldots$ of wave functions that are eigenfunctions for each of them (cf.~Footnote~\ref{fn:33+}). Here we expect that among the $\omega_1,\omega_2,\ldots$ there are groups of many $\omega_n$ on which every macroscopic operator possesses the same eigenvalue, for otherwise carrying out all macroscopically possible observations would allow us to distinguish completely between all of the $\omega_1,\omega_2,\ldots$ (i.e., an absolutely precise determination of the state, which in general is not the case). These groups we denote $\{\omega_{1,p},\ldots,\omega_{s_p,p}\}$, $p=1,2,\ldots$ (replacing the one index $n=1,2,\ldots$ with two indices $p=1,2,\ldots$ and $\lambda=1,\ldots,s_p$)---i.e., the $\omega_{1,p},\ldots,\omega_{s_p,p}$ are degenerate eigenfunctions for all macroscopic quantities.\footnote{A macroscopic quantity is one whose value can \emph{exactly} be determined by means of macroscopic measurements. Thus, if $\AAA$ can assume all values between $-\infty$ and $+\infty$, and if it is characteristic of the macroscopic inaccuracy that only intervals $[k,k+1)$ (for $k=0,\pm1,\pm2,\ldots$) can be distinguished from one another, then only $f(\AAA)$ is macroscopically measurable, with $f$ the following function: $f(x)=k$ for $k\leq x< k+1$ (for $k=0,\pm1,\pm2,\ldots$). Cf., however, the discussion in Section~\ref{sec:2} and Footnote~\ref{fn:+++31}.} Thus, instead of the system $\omega_{1,p},\ldots,\omega_{s_p,p}$, any other system $\omega'_{1,p},\ldots,\omega'_{s_p,p}$ obtained from the former by a unitary transformation would serve the purpose just as well.

If all states of a group $\{\omega_{1,p},\ldots,\omega_{s_p,p}\}$ get mixed with equal weights then one obtains a statistical ensemble with the statistical operator
\be
\frac{1}{s_p} \EEE_p = \frac{1}{s_p} \sum_{\lambda=1}^{s_p} \PPP_{\omega_{\lambda,p}}\,.
\ee
The operator $\EEE_p$ does not change when the $\omega_{\lambda,p}$ get replaced with $\omega'_{\lambda,p}$ just mentioned, as one can easily verify. Every macroscopic operator $\AAA$ has the $\omega_{\lambda,p}$ as eigenfunctions, and thus is a linear combination of the $\PPP_{\omega_{\lambda,p}}$ with the eigenvalues as coefficients,\footnote{A Hermitian operator with eigenfunctions $\chi_1,\chi_2,\ldots$ and respective eigenvalues $w_1,w_2,\ldots$ must be equal to $\sum_n w_n \PPP_{\chi_n}$. See also \cite{vN27b}.} and since all $\omega_{\lambda,p}$ with the same $p$ have the same eigenvalue, $\AAA$ is even a linear combination of the $\EEE_p$, as we note here for future use.

By the way, $\frac{1}{s_p}\EEE_p$ is, as can be seen from the way it arises, the statistical operator of the ensemble in which all macroscopic quantities have the values corresponding to the $p$-th group (where the $s_p$ quantum states have the same weight)---thus, \emph{$\frac{1}{s_p}\EEE_p$ corresponds to the $p$-th one among the alternatives concerning the properties of the system that can be distinguished by macroscopic measurements.} Therefore it is the equivalent of the ``phase cells'' of the Gibbsian statistical mechanics. The number $s_p=\tr \EEE_p$ ($\tr$ means trace, cf.~Footnote~\ref{fn:+++35}) is the number of real (microscopic) states in this cell---its size is therefore a measure of the coarseness of the macroscopic perspective.

\subsection{}\label{sec:I2}

Let us now consider the energy operator $\HHH$ with the eigenfunctions $\varphi_1,\varphi_2,\ldots$ and the eigenvalues $W_1,W_2,\ldots$, so
\be
\HHH=\sum_n W_n \PPP_{\varphi_n}\,.
\ee
We emphasize that $\HHH$ is the \emph{exact} energy and not any macroscopic approximation.

In general, the $\varphi_n$ are different from the $\omega_{\lambda,p}$, and $\HHH$ is not a linear combination of the $\EEE_p$, since the energy is not a macroscopic quantity, as it cannot be measured with absolute precision with macroscopic means.\footnote{For example, think of the situation of observing an ordinary gas. In principle, of course, an energy with point spectrum can, under favorable circumstances, be measured with \emph{absolute} precision: one can, e.g., decide whether an oscillator is in the ground state or not.} With a certain (reduced) accuracy, however, this is indeed possible, so that the energy eigenvalues $W_1,W_2,\ldots$ can be collected in groups $\{W_{1,a},\ldots,W_{S_a,a}\}$, $a=1,2,\ldots$ (again we replace the single index in $W_n$ and $\varphi_n$, $n=1,2,\ldots$, with two indices, $W_{\rho,a}$ and $\varphi_{\rho,a}$ with $a=1,2,\ldots$, $\rho=1,\ldots,S_a$) in such a way that all $W_{\rho,a}$ with the same $a$ are close to each other and only those with different $a$ (i.e., the full groups) can be macroscopically distinguished. How do we formulate the fact that we can macroscopically measure the membership of an energy value in a group $\{W_{1,a},\ldots,W_{S_a,a}\}$?

We do this by means of a trick that we have already mentioned and applied several times in \cite{vN27b}. Let $f_a(x)$ be the function that assumes the value 1 for $x=W_{1,a},\ldots,W_{S_a,a}$ (for fixed $a$!) and is otherwise 0. Thus, $f_a(\HHH)$ is a quantity that has the value 1 when the energy value belongs to the aforementioned group, and is otherwise 0---therefore it can be measured macroscopically. From
\be
\HHH=\sum_n W_n \PPP_{\varphi_n}
\ee
it follows that
\be
f_a(\HHH) = \sum_n f_a(W_n) \PPP_{\varphi_n}
\ee
(cf.~\cite{vN27b}), thus
\be
f_a(\HHH)=\sum_{\rho=1}^{S_a} \PPP_{\varphi_{\rho,a}}\,,
\ee
and this must be a linear combination of the $\EEE_p$. Now the operator $\sum_{\rho=1}^{S_a} \PPP_{\varphi_{\rho,a}}$, and likewise each $\EEE_p=\sum_{\lambda=1}^{s_p}\PPP_{\omega_{\lambda,p}}$, are equal to their own squares, and any two different $\EEE_p$ have product 0\footnote{To prove this, we need to show for two arbitrary but distinct elements $\varphi,\psi$ of an orthogonal system that $\PPP^2_{\varphi}=\PPP_{\varphi}$, $\PPP_{\varphi}\PPP_{\psi}=0$. Let $f$ be any wave function, then we have that (cf.~Section~\ref{sec:3}
\begin{align}
\PPP^2_{\varphi}f&=((f,\varphi)\varphi,\varphi)\varphi= (f,\varphi)(\varphi,\varphi)\varphi=(f,\varphi)\varphi = \PPP_{\varphi}f\,,\\
\PPP_{\varphi}\PPP_{\psi}f&= ((f,\psi)\psi,\varphi)\varphi = (f,\psi)(\psi,\varphi)\varphi=0\,.
\end{align}}---this implies that in the aforementioned linear combination of the $\EEE_p$ each coefficient is equal to its own square, i.e., is either 0 or 1. Thus, $\sum_{\rho=1}^{S_a}\PPP_{\varphi_{\rho,a}}$ is simply the sum of some $\EEE_p$, let them be called $\EEE_{1,a},\ldots,\EEE_{N_a,a}$:
\be
\sum_{\rho=1}^{S_a} \PPP_{\varphi_{\rho,a}} = \sum_{\nu=1}^{N_a} \EEE_{\nu,a}\,.
\ee
By taking the trace, this implies
\be
S_a = \sum_{\nu=1}^{N_a} s_{\nu,a}\,.
\ee
Since the product of
\be
\sum_{\nu=1}^{N_a}\EEE_{\nu,a} \text{ and }
\sum_{\nu=1}^{N_b}\EEE_{\nu,b} \quad (a\neq b)
\ee
is, according to what we said before, equal to the sum of those $\EEE_p$ appearing in both sums, and since, on the other hand, it is also equal to the product of
\be
\sum_{\rho=1}^{N_a}\PPP_{\varphi_{\rho,a}} \text{ and }
\sum_{\rho=1}^{N_b}\PPP_{\varphi_{\rho,b}}\,,
\ee
which vanishes, the sum of the common terms $\EEE_p$ is 0. Therefore there are none, as the sum of several $\EEE_p$, i.e., of several $\PPP_{\omega_{n}}$, never vanishes.\footnote{From $\PPP_{\omega'}+\PPP_{\omega''}+\ldots=0$ (with $\omega'$, $\omega'', \ldots$ pairwise orthogonal) we obtain by multiplication with $\PPP_{\omega'}$ the equation $\PPP_{\omega'}=0$, which is certainly false.} Finally, the $\EEE_{\nu,a}$ exhaust the $\EEE_p$ (so far we have seen merely that they re-index a subset in a one-to-one way); to see this, it suffices to show that
\be
\sum_{a=1}^\infty \sum_{\nu=1}^{N_a} \EEE_{\nu,a} = \sum_{p=1}^\infty \EEE_p\,.
\ee
The left hand side is the sum of all $\EEE_{\nu,a}$, and thus of all $\PPP_{\varphi_{\rho,a}}$, and thus $\1$ (for a complete orthogonal system $\chi_1,\chi_2,\ldots$, the sum of all $\PPP_{\chi_n}$ equals $\1$,\footnote{By inspecting the definition of $\PPP_\chi$ as a matrix in Section~\ref{sec:3} we see that this is identical to the usual form of completeness relation. Cf.\ also \cite{vN27b}.} and the $\varphi_{\rho,a}$ do form a complete orthogonal system); the right hand side is the sum of all $\EEE_p$, and thus of all $\PPP_{\omega_{\lambda,p}}$, and thus $\1$, too (also the $\omega_{\lambda,p}$ form a complete orthogonal system)---thus, everything is proved.

We thus have that the $\EEE_{\nu,a}$ and $s_{\nu,a}$ with $a=1,2,\ldots$, $\nu=1,\ldots,N_a$ is just a different way of indexing the $\EEE_p$ and $s_p$ with $p=1,2,\ldots$. Correspondingly, we write $\omega_{\lambda,\nu,a}$ for $\omega_{\lambda,p}$. We introduce
\be
\DDelta_a = \sum_{\rho=1}^{S_a} \PPP_{\varphi_{\rho,a}} = 
\sum_{\nu=1}^{N_a} \EEE_{\nu,a}\,.
\ee
We see that $\frac{1}{S_a}\DDelta_a$ is the mixture of the states $\varphi_{1,a},\ldots,\varphi_{S_a,a}$ with equal weights, or, alternatively, the mixture of the mixtures $\frac{1}{s_{1,a}}\EEE_{1,a},\ldots, \frac{1}{s_{N_a,a}}\EEE_{N_a,a}$ (considered above as corresponding to phase cells) with weights proportional to $s_{1,a},\ldots,s_{N_a,a}$.

The analoga of these concepts in the Gibbsian theory are, again, obvious: $\frac{1}{S_a}\DDelta_a$ corresponds to the energy surface, i.e., to the micro-canonical ensemble, $N_a$ is the number of phase cells $\EEE_{\nu,a}$ on the energy surface, and $S_a=\tr\DDelta_a$ is the number of true states (i.e., of stationary quantum orbits) on it.

The macroscopically possible energy measurements thus decompose the totality of conceivable states into the energy surfaces $\DDelta_a$, $a=1,2,\ldots$; further energy measurements (which would resolve the $\DDelta_a$ into the $\varphi_{\rho,a}$, $\rho=1,\ldots,S_a$) are not possible with these means. However, other measurements are macroscopically possible, and they must refer to quantities whose operators do not commute with $\HHH$, i.e., which cannot be measured simultaneously with the (microscopic) energy. Classically speaking, they must refer to non-integrals of motion, i.e., to quantities that change with time.\footnote{For example, in a gas enclosed in a box $K$, the total energy of the molecules in the left half of $K$ can be measured macroscopically with certain accuracy---but is not an integral and thus varies with time.} These measurements decompose the energy surface $\DDelta_a$ into the phase cells $\EEE_{\nu,a}$, $\nu=1,\ldots,N_a$. A further decomposition (resolving the $\EEE_{\nu,a}$ into the $\omega_{\lambda,\nu,a}$, $\lambda=1,\ldots,s_{\nu,a}$) is macroscopically impossible.

We thus have that the quantity $N_a$ is a measure of the extent to which the macroscopic methods of measuring are adequate for quantities that cannot simultaneously be measured with energy---i.e., the extent to which the inaccuracy of macroscopic energy measurements is determined by the uncertainty relations. The magnitude of the $s_{\nu,a}$ (i.e., of the phase cells $\EEE_{\nu,a}$), on the other hand, is a measure of the inaccuracy of the macroscopic methods as such, i.e., as a consequence of their imperfection. The inaccuracy due to $N_a$ gets compensated by observations of non-integrals; it is not a weakness of our measurement apparatuses, whereas the inaccuracy due to $s_{\nu,a}$ is. Finally, 
\be
S_a = \sum_{\nu=1}^{N_a} s_{\nu,a}
\ee
is a measure of the product of both, i.e., for the total, actual uncertainty of the energy.

\subsection{}\label{sec:I3}

Suppose now we are given an arbitrary state $\psi$ (where the wave function $\psi$ is normalized, i.e., $\|\psi\|^2 = (\psi,\psi) =1$). The probability that macroscopic measurements on a system in this state will yield the values corresponding to the phase cell $\EEE_{\nu,a}$ is, according to the known rules, the sum of the transition probabilities to the eigenfunctions $\omega_{1,\nu,a},\ldots,\omega_{s_{\nu,a},\nu,a}$ constituting $\EEE_{\nu,a}$. Thus, it is
\be
\sum_{\lambda=1}^{s_{\nu,a}} \bigl| (\psi,\omega_{\lambda,\nu,a}) \bigr|^2
=\sum_{\lambda=1}^{s_{\nu,a}} (\PPP_{\omega_{\lambda,\nu,a}}\psi,\psi)
=(\EEE_{\nu,a}\psi,\psi)\,.
\ee
In words, this is how strongly the cell $\EEE_{\nu,a}$ is occupied in the state $\psi$. Likewise, the probability of the the energy value to belong to the group $\{W_{1,a},\ldots,W_{s_{\nu,a}}\}$ is given by
\be
\sum_{\rho=1}^{S_a}\bigl| (\psi,\varphi_{\rho,a}) \bigr|^2
=\sum_{\rho=1}^{S_a}(\PPP_{\varphi_{\rho,a}}\psi,\psi)
=(\DDelta_a\psi,\psi)\,.
\ee
Thus, it is the occupation number of the energy surface $\DDelta_a$. We note that, in agreement with these concepts,
\begin{align}
\sum_{\nu=1}^{N_a} (\EEE_{\nu,a}\psi,\psi) &= (\DDelta_a \psi,\psi) \\
\sum_{a=1}^\infty (\DDelta_a \psi,\psi) &= (\psi,\psi)=1\,.
\end{align}

Now we are ready to define the micro-canonical ensemble pertaining to the state $\psi$ by specifying its statistical operator. If one $(\DDelta_a \psi,\psi)$ were 1 and the others 0,\footnote{Note that all our ``occupation numbers'' are, by their nature, non-negative.} we would of course have to take the statistical operator $\frac{1}{S_a}\DDelta_a$ considered already in Section~\ref{sec:I2}.\footnote{In \cite{vN27b}, general reasons are provided for the conclusion that always this statistical operator belongs to that ensemble defined by requiring merely that the energy lies in the $a$-th group.} But if several (or all) $(\DDelta_a \psi,\psi)$ are nonzero, we define it to be the mixture of the $\frac{1}{S_1}\DDelta_1,\frac{1}{S_2}\DDelta_2,\ldots$ with weights $(\DDelta_1\psi,\psi), (\DDelta_2\psi,\psi),\ldots$. Thus, the micro-canonical ensemble has the statistical operator
\be
\UUU_\psi = \sum_{a=1}^\infty \frac{(\DDelta_a \psi,\psi)}{S_a} \DDelta_a\,.
\ee
Of course, this definition is really justified only afterwards by its success, i.e., by the fact that only with this definition, the ergodic theorem and the $H$-theorem hold. (Practically, of course, all but one $(\DDelta_a\psi,\psi)$ are very small.)

It remains to define the entropies of $\psi$ and $\UUU_\psi$ (of the state and of the corresponding (virtual) micro-canonical ensemble). The expressions for entropy given by the author in \cite{vN27c} are not applicable here in the way they were intended, as they were computed from the perspective of an observer who can carry out all measurements that are possible in principle---i.e., regardless of whether they are macroscopic (for example, there every pure state has entropy 0, only mixtures have entropies greater than 0!). If we take into account that the observer can measure only macroscopically then we find different entropy values (in fact, greater ones, as the observer is now less skilful and possibly can therefore extract less mechanical work from the system); nevertheless, the theory can be set up also in this case. How to do this has been discussed by E.~Wigner,\footnote{Mr.~E.~Wigner has communicated his hitherto unpublished results on this topic to the author orally. Here we shall use only those formulas necessary for the purpose at hand, while we need not enter into the general theory.} the formulas for the entropies $S(\psi), S(\UUU_\psi)$ of $\psi$ and $\UUU_\psi$ read:\footnote{We have omitted the usual factor $k$ (= Boltzmann constant), and thus introduced as the unit of temperature ``erg'' per degree of freedom. [1 erg = 1 g cm$^2$/s$^2$ = $10^{-7}$ J]}
\begin{align}
S(\psi)&=
-\sum_{a=1}^\infty \sum_{\nu=1}^{N_a} (\EEE_{\nu,a}\psi,\psi) \ln \frac{(\EEE_{\nu,a}\psi,\psi)}{s_{\nu,a}}\,,\\
S(\UUU_\psi)&=
-\sum_{a=1}^\infty (\DDelta_a\psi,\psi) \ln \frac{(\DDelta_a \psi,\psi)}{S_a}\,.
\end{align}
By the way, these entropy formulas are identical to the usual ones based on Boltzmann's definition of entropy (and Stirling's formula), as one sees by noting that the $(\EEE_{\nu,a}\psi,\psi)$ (the $(\DDelta_a\psi,\psi)$) are the relative occupation numbers of the phase cells (of the energy surfaces) and the $s_{\nu,a}$ (the $S_a$) are the numbers of quantum orbits therein, i.e., their so-called a-priori weights.

\section{Implementation of Proofs}\label{sec:II}

\subsection{}\label{sec:II1}

The temporal evolution $\psi_t$ of the initial state $\psi$ is determined by the time-dependent Schr\"odinger differential equation
\be
\psi_0=\psi\,,\quad
\frac{\partial}{\partial t}\psi_t = \frac{i}{\hbar} \HHH \psi_t
\ee
with $\HHH$ the energy operator,
\be
\HHH = \sum_{a=1}^\infty \sum_{\rho=1}^{S_a} W_{\rho,a} \PPP_{\rho,a}\,.
\ee
Thus, if
\be
\psi = \sum_{a=1}^\infty \sum_{\rho=1}^{S_a} r_{\rho,a} e^{i\alpha_{\rho,a}}\varphi_{\rho,a}
\ee
with $r_{\rho,a}\geq 0$ and $0\leq \alpha_{\rho,a}< 2\pi$ then
\be
\psi_t = \sum_{a=1}^\infty \sum_{\rho=1}^{S_a} r_{\rho,a} e^{i(W_{\rho,a} t/\hbar + \alpha_{\rho,a})}\varphi_{\rho,a}\,.
\ee
We introduce the abbreviations
\be
x_{\nu,a} = (\EEE_{\nu,a} \psi_t,\psi_t)\,,\quad
u_a = (\DDelta_a \psi_t,\psi_t) = (\DDelta_a\psi,\psi)
\ee
(the last two expressions are equal because
\be
(\DDelta_a\psi_t,\psi_t) =\sum_{\rho=1}^{S_a} (\PPP_{\varphi_{\rho,a}} \psi_t,\psi_t)
=\sum_{\rho=1}^{S_a} \bigl| (\psi_t,\varphi_{\rho,a}) \bigr|^2
=\sum_{\rho=1}^{S_a} r_{\rho,a}^2
\ee
does not depend on $t$.) As we see, 
\be
\sum_{\nu=1}^{N_a} x_{\nu,a} = u_a\,,
\ee
\be
\sum_{a=1}^\infty u_a= 1\,,
\ee
$x_{\nu,a}$ depends on $t$, $u_a$ does not.\footnote{Thus, the micro-canonical ensemble [i.e., density matrix] $\UUU_{\psi}= \sum_{a=1}^\infty (u_a/S_a)\DDelta_a$ does not change when $\psi$ is replaced with $\psi_t$.} From [our discussion above at] the definitions of entropies we know that the $x_{\nu,a},u_a$ are non-negative and that
\be
S(\psi_t) = -\sum_{a=1}^\infty \sum_{\nu=1}^{N_a} x_{\nu,a} \ln \frac{x_{\nu,a}}{s_{\nu,a}}\,,\quad
S(\UUU_{\psi}) = -\sum_{a=1}^\infty u_a \ln \frac{u_a}{S_a}\,.
\ee
Since the sum of all $x_{\nu,a}$ (or of all $u_a$) equals 1, they all lie in $[0,1]$, and thus both entropies are always non-negative. We now discuss more closely their magnitudes.

We note $0\leq x_{\nu,a}\leq u_a$; we replace $x_{\nu,a}$ by a variable $z$ and assume first that
\be
0\leq z \leq \frac{2s_{\nu,a}}{S_a} u_a\,,\quad
\text{i.e., } \Bigl|\frac{S_a}{s_{\nu,a}u_a}z-1\Bigr| \leq 1\,.
\ee 
Then
\be
-z\ln\frac{z}{s_{\nu,a}} = -\frac{s_{\nu,a}u_a}{S_a}\Bigl(1+ \Bigl[ \frac{S_a}{s_{\nu,a}u_a}z-1\Bigr]\Bigr)\biggr(\ln \frac{u_a}{S_a} 
+\ln\Bigl(1+\Bigl[\frac{S_a}{s_{\nu,a}u_a}z-1\Bigr]\Bigr)\biggr)
\ee
\begin{multline}
= -\frac{s_{\nu,a}u_a}{S_a}\Bigl(1+\Bigl[\frac{S_a}{s_{\nu,a}u_a}z-1\Bigr]\Bigr)
\biggl(\ln\frac{u_a}{S_a} + \Bigl[ \frac{S_a}{s_{\nu,a}u_a}z-1\Bigr]\\
-\frac{1}{2}\Bigl[ \frac{S_a}{s_{\nu,a}u_a}z-1 \Bigr]^2+
\frac{1}{3}\Bigl[ \frac{S_a}{s_{\nu,a}u_a}z-1  \Bigr]^3-+\ldots\biggr)
\end{multline}
\begin{multline}
=-\frac{s_{\nu,a}u_a}{S_a}\ln\frac{u_a}{S_a}- \frac{s_{\nu,a}u_a}{S_a} \Bigl( \ln\frac{u_a}{S_a}+1\Bigr)\Bigl[\frac{S_a}{s_{\nu,a}u_a}z-1 \Bigr]\\
-\frac{s_{\nu,a}u_a}{1\times2 S_a}\Bigl[\frac{S_a}{s_{\nu,a}u_a}z-1 \Bigr]^2
+\frac{s_{\nu,a}u_a}{2\times 3 S_a}\Bigl[\frac{S_a}{s_{\nu,a}u_a}z-1 \Bigr]^3-+\ldots\,.
\end{multline}
Since
\be
\frac{1}{1\times 2} + \frac{1}{2\times 3}+ \ldots =1\,,
\ee
the sum of the absolute values of the last terms is no greater than
\be
\frac{s_{\nu,a}u_a}{S_a}\Bigl[\frac{S_a}{s_{\nu,a}u_a}z-1 \Bigr]^2\,,
\ee
and we can thus write
\be
\label{eq1}
\biggl| -\frac{s_{\nu,a}}{S_a} u_a \ln \frac{u_a}{S_a} - \Bigl(\ln\frac{u_a}{S_a}+1  \Bigr) \Bigl[ z-\frac{s_{\nu,a}}{S_a}u_a \Bigr] +z\ln \frac{z}{s_{\nu,a}}  \biggr|
\leq \frac{S_a}{s_{\nu,a} u_a}\Bigl[z-\frac{s_{\nu,a}u_a}{S_a} \Bigr]^2\,.
\ee
In order to prove this also for the other values of $z$, we compare the left hand side (without $|\cdots|$) with half of the right hand side. For $z=s_{\nu,a}u_a/S_a$ they both vanish, and their derivatives are in general
\be
-\Bigl( \ln\frac{u_a}{S_a} +1 \Bigr) + \Bigl( \ln \frac{z}{s_{\nu,a}}+1\Bigr) =
\ln \frac{S_a}{s_{\nu,a}u_a}z
\ee
and
\be
\frac{S_a}{s_{\nu,a}u_a} \Bigl[ z-\frac{s_{\nu,a}u_a}{S_a} \Bigr]=
\frac{S_a}{s_{\nu,a}u_a}z-1\,.
\ee
Obviously, the former is always less than or equal to the latter and $\gtreqqless 0$ when
\be\label{cond1}
z\gtreqqless \frac{s_{\nu,a}u_a}{S_a}\,.
\ee
Thus, the left hand side of \eqref{eq1}, while always non-negative, is $\gtreqqless$ than half the right hand side of \eqref{eq1} for $z$ as in \eqref{cond1}. We thus have in general that
\be
0\leq -\frac{s_{\nu,a}}{S_a}u_a \ln \frac{u_a}{S_a} - \Bigl( \ln\frac{u_a}{S_a} + 1 \Bigr) \Bigl[z-\frac{s_{\nu,a}}{S_a}u_a\Bigr] + z \ln \frac{z}{s_{\nu,a}u_a} 
\leq \frac{S_a}{s_{\nu,a}u_a}\Bigl[z-\frac{s_{\nu,a}u_a}{S_a}\Bigr]^2.
\ee
Now we set $z=x_{\nu,a}$ and sum over $\nu=1,\ldots,N_a$; since
\be
\sum_{\nu=1}^{N_a} s_{\nu,a} = S_a\,,\quad
\sum_{\nu=1}^{N_a} x_{\nu,a}=u_a\,,
\ee
we obtain that
\be
0 \leq -u_a\ln\frac{u_a}{S_a} + \sum_{\nu=1}^{N_a} x_{\nu,a}\ln\frac{x_{\nu,a}}{s_{\nu,a}}
\leq \sum_{\nu=1}^{N_a} \frac{S_a}{s_{\nu,a}u_a}\Bigl[ x_{\nu,a}-\frac{s_{\nu,a}u_a}{S_a} \Bigr]^2.
\ee
If we sum also over $a=1,2,\ldots,$ we obtain that
\be
0\leq S(\UUU_\psi)-S(\psi_t) 
\leq \sum_{a=1}^\infty \sum_{\nu=1}^{N_a} \frac{S_a}{s_{\nu,a}u_a}\Bigl[x_{\nu,a}-\frac{s_{\nu,a}u_a}{S_a} \Bigr]^2.
\ee

This estimate provides an ansatz for proving the $H$-theorem. We now proceed to the ergodic theorem and find that it requires a bound on the same expression.

\subsection{}\label{sec:II2}

Let $\AAA$ be a macroscopically observable quantity, i.e.,
\be
\AAA = \sum_{a=1}^\infty \sum_{\nu=1}^{N_a} \eta_{\nu,a}\EEE_{\nu,a}\,.
\ee
The $\omega_{\lambda,\nu,a}$ of the phase cell $\EEE_{\nu,a}$ are eigenfunctions of $\AAA$ with eigenvalue $\eta_{\nu,a}$---i.e., $\eta_{\nu,a}$ is the value of $\AAA$ in the phase cell $\EEE_{\nu,a}$. Thus, $\AAA$ has the following expectation values in the state $\psi_t$ and in the micro-canonical ensemble $\UUU_\psi$:
\begin{align}
(\AAA\psi_t,\psi_t) &= \sum_{a=1}^\infty \sum_{\nu=1}^{N_a} \eta_{\nu,a} (\EEE_{\nu,a}\psi_t,\psi_t) = \sum_{a=1}^\infty \sum_{\nu=1}^{N_a} \eta_{\nu,a} x_{\nu,a}\,,
\label{AAAexpectpsit}\\
\tr(\AAA\UUU_\psi) &=
\tr\biggl(\Bigl(\sum_{a=1}^\infty\sum_{\nu=1}^{N_a} \eta_{\nu,a} \EEE_{\nu,a} \Bigr)\Bigl(\sum_{a=1}^\infty \sum_{\nu=1}^{N_a} \frac{u_a}{S_a}\EEE_{\nu,a} \Bigr) \biggr)\\
&= \sum_{a=1}^\infty\sum_{\nu=1}^{N_a}\eta_{\nu,a} \frac{s_{\nu,a}u_a}{S_a}\,.
\label{AAAexpectUUU}
\end{align}
(The number of terms gets reduced by the fact that $\EEE_{\nu,a}\EEE_{\mu,b}=0$ except when $\nu=\mu$ and $a=b$, in which case $\EEE_{\nu,a}\EEE_{\mu,b}=\EEE_{\nu,a}$ has the trace $u_{\nu,a}$.) We denote the values \eqref{AAAexpectpsit} and \eqref{AAAexpectUUU} by $E_{\AAA}(\psi_t)$ and $E_{\AAA}(\UUU_\psi)$. Using the Schwarz inequality, we can estimate:
\begin{align}
&\bigl( E_{\AAA}(\psi_t) - E_{\AAA}(\UUU_\psi) \bigr)^2=
\biggl( \sum_{a=1}^\infty\sum_{\nu=1}^{N_a} \eta_{\nu,a} \Bigl[ x_{\nu,a} - \frac{s_{\nu,a}u_a}{S_a}\Bigr] \biggr)^2\\
&= \biggl(\sum_{a=1}^\infty \sum_{\nu=1}^{N_a} \sqrt{\frac{s_{\nu,a}u_a}{S_a}} \eta_{\nu,a} \sqrt{\frac{S_a}{s_{\nu,a}u_a}} \Bigl[ x_{\nu,a} - \frac{s_{\nu,a}u_a}{S_a}\Bigr]  \biggr)^2\\
&\leq \Bigl(\sum_{a=1}^\infty \sum_{\nu=1}^{N_a} \frac{s_{\nu,a}u_a}{S_a} \eta_{\nu,a}^2 \Bigr) \Bigl( \sum_{a=1}^\infty \sum_{\nu=1}^{N_a} \frac{S_a}{s_{\nu,a}u_a}\Bigl[x_{\nu,a}-\frac{s_{\nu,a}u_a}{S_a} \Bigr]^2\Bigr)\,.
\end{align}
The first factor we abbreviate $\bar{\eta}^2$; since
\begin{align}
\frac{s_{\nu,a}u_a}{S_a} &\geq 0\,,\\
\sum_{a=1}^\infty \sum_{\nu=1}^{N_a} \frac{s_{\nu,a}u_a}{S_a} &= 1\,,\\
\sum_{a=1}^\infty\sum_{\nu=1}^{N_a} \frac{s_{\nu,a}u_a}{S_a} \eta_{\nu,a}^2 &= \bar{\eta}^2\,,
\end{align}
this is a weighted average of the values $\eta_{\nu,a}^2$ of $\AAA^2$, in fact the micro-canonical average: after all, $\UUU_\psi$ is the mixture of the $(1/S_a) \DDelta_a$ ($a=1,2,\ldots$) with weights $u_a$ and thus that of the $(1/s_{\nu,a})\EEE_{\nu,a}$ ($a=1,2,\ldots$; $\nu=1,\ldots,N_a$) with weights $s_{\nu,a}u_a/S_a$, and $\AAA^2$ has, as we know, the value $\eta_{\nu,a}^2$ in $(1/s_{\nu,a}) \EEE_{\nu,a}$. Thus, $\bar{\eta}$ is a reasonable measure of the order of magnitude of the quantity $\AAA$. We thus have that
\be
\bigl( E_{\AAA}(\psi_t) - E_{\AAA}(\UUU_\psi) \bigr)^2
\leq \bar{\eta}^2 \sum_{a=1}^\infty\sum_{\nu=1}^{N_a} \frac{S_a}{s_{\nu,a}u_a} \Bigl[ s_{\nu,a} - \frac{s_{\nu,a}u_a}{S_a}\Bigr]^2.
\ee

\subsection{}\label{sec:II3}

Now we average over time, denoted by $M_t$. We thus obtain that
\begin{align}
M_t \bigl\{ |S(\UUU_\psi) - S(\psi_t)| \bigr\} 
&\leq M_t \biggl\{ \sum_{a=1}^\infty \sum_{\nu=1}^{N_a} \frac{S_a}{s_{\nu,a}u_a}\Bigl[x_{\nu,a}-\frac{s_{\nu,a}u_a}{S_a} \Bigr]^2 \biggr\},\\
M_t \bigl\{\bigl( E_{\AAA}(\psi_t) - E_{\AAA}(\UUU_\psi) \bigr)^2\bigr\}
&\leq \bar{\eta}^2 M_t 
\biggl\{ \sum_{a=1}^\infty \sum_{\nu=1}^{N_a} \frac{S_a}{s_{\nu,a}u_a}\Bigl[x_{\nu,a}-\frac{s_{\nu,a}u_a}{S_a} \Bigr]^2 \biggr\}.
\end{align}
Thus, ergodic theorem and $H$-theorem will both be established when we have shown that the $M_t\{\cdots\}$ on the right hand side is small uniformly for all initial states $\psi$ (i.e., all $r_{\rho,a},\alpha_{\rho,a}$ with $\sum_{a=1}^\infty \sum_{\rho=1}^{S_a} r^2_{\rho,a}=\|\psi\|^2=1$). (Note that while $x_{\nu,a}$ depends on $t$, $r_{\rho,a}$, and $\alpha_{\rho,a}$, the $u_a$ depend only on $r_{\rho,a}$, and everything else is constant.)

In order to show this we first compute $x_{\nu,a}$:\footnote{The number of terms gets reduced by the fact that $(\EEE_{\nu,a}\varphi_{\rho,b},\varphi_{\sigma,c})=(\varphi_{\rho,b},\EEE_{\nu,a}\varphi_{\sigma,c})=0$ unless $a=b=c$. It suffices to show $\EEE_{\nu,a}\varphi_{\rho,b}=0$ for $a\neq b$, or (because of $\EEE_{\nu,a}\DDelta_a = \EEE_{\nu,a}$, see Section~\ref{sec:I2}) that $\DDelta_a \varphi_{\rho,b}=0$. This follows from $\DDelta_a = \sum_{\sigma=1}^{S_a} \PPP_{\varphi_{\sigma,a}}$, since $\varphi_{\rho,b}$ is orthogonal to all $\varphi_{\sigma,a}$.}
\begin{align}
x_{\nu,a} &= (\EEE_{\nu,a}\psi_t,\psi_t)\\
&= \biggl( \sum_{b=1}^\infty \sum_{\rho=1}^{S_b} r_{\rho,b} e^{i(W_{\rho,b}t/\hbar+\alpha_{\rho,b})}\EEE_{\nu,a}\varphi_{\rho,b},\nonumber\\
&\quad \sum_{b=1}^\infty \sum_{\rho=1}^{S_b} r_{\rho,b} e^{i(W_{\rho,b}t/\hbar+\alpha_{\rho,b})}\varphi_{\rho,b} \biggr)\\
&=\sum_{\rho,\sigma=1}^{S_a} r_{\rho,a} r_{\sigma,a} e^{i\bigl((W_{\rho,a}-W_{\sigma,a})t/\hbar+(\alpha_{\rho,a}-\alpha_{\sigma,a})\bigr)} (\EEE_{\nu,a}\varphi_{\rho,a}, \varphi_{\sigma,a})\,.
\end{align}
Thus, using $\sum\limits_{\rho=1}^{S_a} r_{\rho,a}^2 = u_a$,
\begin{multline}
x_{\nu,a}-\frac{s_{\nu,a}u_a}{S_a}=
\sum_{\substack{\rho,\sigma=1\\\rho\neq\sigma}}^{S_a} r_{\rho,a}r_{\sigma,a} e^{i\bigl((W_{\rho,a}-W_{\sigma,a})t/\hbar + (\alpha_{\rho,a}-\alpha_{\sigma,a})  \bigr)} \times\\
\times (\EEE_{\nu,a}\varphi_{\rho,a},\varphi_{\sigma,a}) +
\sum_{\rho=1}^{S_a} r_{\rho,a}^2 \Bigl\{ (\EEE_{\nu,a}\varphi_{\rho,a},\varphi_{\rho,a})-\frac{s_{\nu,a}}{S_a} \Bigr\}\,.
\end{multline}
If we square this expression and average it over $t$ then all terms containing $e^{ict}$ with $c\neq 0$ vanish. Thus, if
\begin{align}
\text{for }\rho\neq\sigma:\quad & W_\rho-W_\sigma \neq 0\,,\label{nondeg}\\
\text{for }\rho\neq\sigma\,,\:\rho'\neq\sigma':\quad &
(W_\rho-W_\sigma)-(W_{\rho'}-W_{\sigma'})\neq 0 \label{nores}
\end{align}
unless $\rho=\rho'$, $\sigma=\sigma'$---i.e., if for every fixed $a$ all $W_{\rho,a}$ ($\rho=1,2,\ldots$) are distinct, and so are all $W_{\rho,a}-W_{\sigma,a}$ ($\rho\neq\sigma$, $\rho,\sigma=1,2,\ldots$)---then we obtain that
\begin{align}
M_t \Bigl( \Bigl[x_{\nu,a}-\frac{s_{\nu,a}u_a}{S_a} \Bigr]^2\Bigr) 
&= \sum_{\substack{\rho,\sigma=1\\\rho\neq\sigma}}^{S_a} r_{\rho,a}^2 r_{\sigma,a}^2 \bigl|(\EEE_{\nu,a}\varphi_{\rho,a},\varphi_{\sigma,a}) \bigr|^2\nonumber\\
&\quad + \biggl( \sum_{\rho=1}^{S_a} r_{\rho,a}^2 \Bigl\{(\EEE_{\nu,a}\varphi_{\rho,a},\varphi_{\rho,a})-\frac{s_{\nu,a}}{S_a} \Bigr\}\biggr)^2.
\end{align}
We now set
\begin{align}
\max_{\substack{\rho,\sigma=1\\\rho\neq\sigma}}^{S_a} \bigl( \bigl| (\EEE_{\nu,a}\varphi_{\rho,a},\varphi_{\sigma,a}) \bigr|^2 \bigr) 
&= \Mb_{\nu,a}\,,\\
\max_{\rho=1}^{S_a} \Bigl( \Bigl\{ (\EEE_{\nu,a}\varphi_{\rho,a},\varphi_{\rho,a}) - \frac{s_{\nu,a}}{S_a}\Bigr\}^2 \Bigr) 
&= \Nb_{\nu,a}\,,
\end{align}
where $\Mb_{\nu,a}$, $\Nb_{\nu,a}$ are constants, i.e., independent of $t$, $r_{\rho,a}$, $\alpha_{\rho,a}$, and thus of $\psi_t$. Since $\sum\limits_{\rho=1}^{S_a} r_{\rho,a}^2 = u_a$, we have that
\begin{align}
M_t \Bigl( \Bigl[x_{\nu,a}-\frac{s_{\nu,a}u_a}{S_a} \Bigr]^2\Bigr) 
&= \sum_{\substack{\rho,\sigma=1\\\rho\neq\sigma}}^{S_a} r_{\rho,a}^2 r_{\sigma,a}^2 \Mb_{\nu,a}
+ \Bigl(\sum_{\rho=1}^{S_a} r_{\rho,a}^2 \sqrt{\Nb_{\nu,a}} \Bigr)^2\\
&\leq u_a^2 (\Mb_{\nu,a} + \Nb_{\nu,a})\,,
\end{align}
and thus
\be
M_t \biggl\{\sum_{a=1}^\infty \sum_{\nu=1}^{N_a} \frac{S_a}{s_{\nu,a}u_a} \Bigl[x_{\nu,a}-\frac{s_{\nu,a}u_a}{S_a} \Bigr]^2\biggr\} 
\leq \sum_{a=1}^\infty \sum_{\nu=1}^{N_a}\frac{S_au_a}{s_{\nu,a}} (\Mb_{\nu,a}+\Nb_{\nu,a})\,.
\ee
Because of $\sum\limits_{a=1}^\infty u_a =1$, this is
\be
\leq \max_{a=1,2,\ldots} \sum_{\nu=1}^{N_a} \frac{S_a}{s_{\nu,a}} (\Mb_{\nu,a}+\Nb_{\nu,a})\,,
\ee
where it suffices to take the maximum over those $a$ for which $u_a\neq 0$, i.e., whose energy surfaces actually occur in the micro-canonical ensemble. Thus, we will have reached our goal when we can prove, for these $a$, that
\be\label{expr1}
\sum_{\nu=1}^{N_a} \frac{S_a}{s_{\nu,a}} (\Mb_{\nu,a}+\Nb_{\nu,a})
\ee
is small; in fact, our result will then hold for all of these $\psi$, as the expression \eqref{expr1} is constant, i.e., independent of $\psi$ (and $t$, $r_{\rho,a}$, $\alpha_{\rho,a}$)---it only involves the $\EEE_{\nu,a}$ (and thus indirectly $S_a$, $N_a$, $s_{\nu,a}$, $\DDelta_a$, and the $\omega_{\lambda,\nu,a}$). In order to bound the expression \eqref{expr1}, we need to bound the $\Mb_{\nu,a}$ and $\Nb_{\nu,a}$.

\subsection{}\label{sec:II4}

We regard $\HHH$ (and thus the $W_{\rho,a}$ and $\varphi_{\rho,a}$) as fixed (obeying \eqref{nondeg} and \eqref{nores}\footnote{These conditions could be relaxed slightly. We could dispense with [\eqref{nondeg}, i.e.,] the distinctness of the $W_{\rho,a}$ and demand the following of $W_{\rho,a}-W_{\sigma,a}$ [instead of \eqref{nores}]: it be possible to partition the set of all pairs $\rho,\sigma$ with $\rho\neq \sigma$ (where $\rho,\sigma=1,\ldots,S_a$) into $k$ groups in such a way that within each group the $W_{\rho,a}-W_{\sigma,a}$ are pairwise distinct---if $k$ is a fixed number for each $a$ and the conditions on the size of the $S_a$, $N_a$, and $s_{\nu,a}$ that we will specify later are satisfied to a sufficient extent then our conclusion is not affected. That is, it does no harm if our conditions \eqref{nondeg} and \eqref{nores} are violated in few cases. We do not give further detail. (In particular, to drop \eqref{nondeg} does not gain us much, as $W_{\rho,a}=W_{\sigma,a}$ and $W_{\rho',a}=W_{\sigma',a}$ together imply that $W_{\rho,a}-W_{\sigma,a} = W_{\rho',a}-W_{\sigma',a}$.)}), as well as the $S_a$, $N_a$, $s_{\nu,a}$, and $\DDelta_a$; we merely vary the $\EEE_{\nu,a}$, within these boundaries. That is, we vary the orthogonal system $\omega_{\lambda,\nu,a}$ ($\nu=1,\ldots,N_a$; $\lambda=1,\ldots,s_{\nu,a}$), subject only to the condition
\be
\sum_{\nu=1}^{N_a} \sum_{\lambda=1}^{s_{\nu,a}} \PPP_{\omega_{\lambda,\nu,a}}=\DDelta_a\,,
\ee
and set
\be
\EEE_{\nu,a} =\sum_{\lambda=1}^{s_{\nu,a}} \PPP_{\omega_{\lambda,\nu,a}}
\ee
for $\nu=1,\ldots,N_a$. Note that all such orthogonal systems $\omega_{\lambda,\nu,a}$ arise from one of them, say $\overline{\omega}_{\lambda,\nu,a}$, by unitary transformations (in $\sum_{\nu=1}^{N_a} s_{\nu,a} = S_a$ dimensions since we keep $a$ fixed). (Think, for example, of the definition of the $\PPP_{\omega}$ as matrices in Section~\ref{sec:3}.)

Then the $\Mb_{\nu,a}$ and $\Nb_{\nu,a}$ depend only on the $\omega_{\lambda,\nu,a}$; not for every choice of the latter, in fact, they are as small as we need them to be (and no reasonable condition on $S_a$, $N_a$, $s_{\nu,a}$ would help with this). For example, if the $\omega_{\lambda,\nu,a}$ coincide with the $\varphi_{\rho,a}$ (where $a$ is fixed, note that there are $S_a$ of each), one sees that every $(\EEE_{\nu,a}\varphi_{\rho,a},\varphi_{\rho,a})$ assumes [for some $\rho$] the value 1 among others, and therefore
\be
\Nb_{\nu,a}\geq \Bigl( 1-\frac{s_{\nu,a}}{S_a}\Bigr)^2 \geq \frac{1}{4}
\ee
(provided that, as is always the case, $s_{\nu,a}\leq \tfrac{1}{2}S_a$ for all $\nu$), and therefore
\be
\sum_{\nu=1}^{N_a} \frac{S_a}{s_{\nu,a}}(\Mb_{\nu,a}+\Nb_{\nu,a}) \geq N_a \times 2\times \frac{1}{4} = \frac{N_a}{2}\,,
\ee
thus arbitrarily large if $N_a$ is large. The unfavorable result in this case arises, of course, from the fact that this choice of $\omega_{\lambda,\nu,a}$ does not represent well their physical meaning: here, the $\EEE_{\nu,a}$ have the same eigenfunctions as $\HHH$ and thus commute with $\HHH$---which we expected not to be the case (cf.\ Section~\ref{sec:I2})!

On the other hand, this behavior is singular and exceptional, and for the overwhelming majority of the relevant systems $\omega_{\lambda,\nu,a}$ we find the right order of magnitude for $\Mb_{\nu,a}$ and $\Nb_{\nu,a}$. But before we prove this, we would like to get an idea (in an inexact way!) of what to expect of $\Mb_{\nu,a}$ and $\Nb_{\nu,a}$ in the best case. To this end we proceed as follows. Instead of averaging
\begin{align}
\Mb_{\nu,a} &= \max_{\substack{\rho,\sigma=1\\\rho\neq \sigma}}^{S_a}\Bigl(
\bigl|(\EEE_{\nu,a}\varphi_{\rho,a},\varphi_{\sigma,a})\bigr|^2 \Bigr)\,,\\
\Nb_{\nu,a} &= \max_{\rho=1}^{S_a} \Bigl( \Bigl\{ (\EEE_{\nu,a}\varphi_{\rho,a},\varphi_{\rho,a}) - \frac{s_{\nu,a}}{S_a}\Bigr\}^2 \Bigr)
\end{align}
over all possible systems $\omega_{\lambda,\nu,a}$ (i.e., of determining which values are predominantly assumed; the definition of the averaging procedure will be explained in the appendix; see also the discussion in Section~\ref{sec:III1}), we average the
\begin{align}
 \bigl| (\EEE_{\nu,a}\varphi_{\rho,a},\varphi_{\sigma,a}) \bigr|^2
&\quad(\rho\neq \sigma, \rho,\sigma=1,\ldots,S_a)\\
\Bigl\{ (\EEE_{\nu,a}\varphi_{\rho,a},\varphi_{\rho,a}) -\frac{s_{\nu,a}}{S_a}\Bigr\}^2
&\quad (\rho=1,\ldots,S_a)
\end{align}
themselves and then take the maximum. That is, we replace the mean of the maximum by the maximum of the mean---this leads to wrong, in fact too small (i.e., too favorable) numbers, but may suffice for the purpose of a first orientation.

As will be shown in the appendix, the averages of
\be
\bigl|(\EEE_{\nu,a}\varphi_{\rho,a},\varphi_{\sigma,a})\bigr|^2 \:\:(\rho\neq\sigma),\quad
(\EEE_{\nu,a}\varphi_{\rho,a},\varphi_{\rho,a})\,, \quad
\Bigl\{(\EEE_{\nu,a}\varphi_{\rho,a},\varphi_{\rho,a})-\frac{s_{\nu,a}}{S_a} \Bigr\}^2
\ee
are equal to, respectively,
\be
\frac{s_{\nu,a}(S_a-s_{\nu,a})}{S_a(S_a^2-1)}\,,\quad
\frac{s_{\nu,a}}{S_a}\,,\quad
\frac{s_{\nu,a}(S_a-s_{\nu,a})}{S_a^2(S_a+1)}\,,
\ee
and thus, if (as is the case in practice) $s_{\nu,a}\ll S_a$, approximately equal to, respectively,
\be
\frac{s_{\nu,a}}{S_a^2}\,,\quad
\frac{s_{\nu,a}}{S_a}\,,\quad
\frac{s_{\nu,a}}{S_a^2}\,.
\ee
For $\Mb_{\nu,a}$, $\Nb_{\nu,a}$ we tentatively insert $s_{\nu,a}/S_a^2$, which yields
\be\label{orientation}
\sum_{\nu=1}^{N_a} \frac{S_a}{s_{\nu,a}} (\Mb_{\nu,a}+\Nb_{\nu,a}) 
= 2 \sum_{\nu=1}^{N_a} \frac{1}{S_a} = \frac{2N_a}{S_a}\,.
\ee
This is small when $N_a/S_a$ is small, i.e., when
\be
\frac{\sum\limits_{\nu=1}^{N_a} s_{\nu,a}}{N_a} = \frac{S_a}{N_a}
\ee 
is large. That is, the $s_{\nu,a}$ (i.e., the phase cells) must be large on average. This result is very reasonable, and we thus proceed to considering the correct average of $\Mb_{\nu,a}$, $\Nb_{\nu,a}$ over the $\omega_{\lambda,\nu,a}$.

\subsection{}\label{sec:II5}

For the average of $\Mb_{\nu,a}$, $\Nb_{\nu,a}$ over all $\omega_{\lambda,\nu,a}$ with
\be
\sum_{\nu=1}^{N_a} \sum_{\lambda=1}^{s_{\nu,a}} \PPP_{\omega_{\lambda,\nu,a}} = \DDelta_a
\ee
we will find in the appendix the respective upper bounds
\be
\frac{\ln S_a}{S_a}\,, \quad
\frac{9s_{\nu,a}\ln S_a}{S_a^2}\,.
\ee
We see that they are $S_a \ln S_a/s_{\nu,a}$ times (respectively $9\ln S_a$ times) larger than the values used in \eqref{orientation} (keep in mind $1\ll s_{\nu,a} \ll S_a$); in particular, the first bound is much worse than the second. It is possible that our bounds can be improved considerably and can get closer to the values of the previous section---we emphasize this so that readers get the right picture of the conditions on the sizes of $S_a,N_a$, and $s_{\nu,a}$ that we will find: they are certainly sufficient but perhaps not necessary.

By inserting the above expressions, we find the average of
\be
\sum_{\nu=1}^{N_a} \frac{S_a}{s_{\nu,a}} (\Mb_{\nu,a}+\Nb_{\nu,a})
\ee
to be
\be\label{expr}
\leq \sum_{\nu=1}^{N_a} \frac{S_a}{s_{\nu,a}} \Bigl( \frac{\ln S_a}{S_a} + \frac{9s_{\nu,a}\ln S_a}{S_a^2} \Bigr) = (\ln S_a)\Bigl( \frac{9N_a}{S_a} + \sum_{\nu=1}^{N_a} \frac{1}{s_{\nu,a}} \Bigr)\,.
\ee
We introduce the arithmetic and the harmonic mean of the $s_{\nu,a}$ ($\nu=1,\ldots,N_a$):
\be
\bar{s}_a = \frac{1}{N_a} \sum_{\nu=1}^{N_a} s_{\nu,a} = \frac{S_a}{N_a}\,,\quad
\frac{1}{\bar{\bar{s}}_a} = \frac{1}{N_a} \sum_{\nu=1}^{N_a} \frac{1}{s_{\nu,a}}\,.
\ee
Then the expression \eqref{expr} equals
\be
(\ln S_a) \Bigl( \frac{9}{\bar{s}_a} + \frac{N_a}{\bar{\bar{s}}_a}\Bigr)\,.
\ee
Because of $\bar{\bar{s}}_a \leq \bar{s}_a$ and $N_a \gg 1$ (which amounts to the justified assumption that the energy surface contains many phase cells), this is approximately equal to
\be\label{expr5}
(\ln S_a)\frac{N_a}{\bar{\bar{s}}_a}\,.
\ee
When is this expression small?

Certainly we must have that $\bar{s}_a \geq \bar{\bar{s}}_a \gg N_a$ and thus $\ln \bar{s}_a \geq \ln N_a$, so we can replace $\ln S_a = \ln \bar{s}_a + \ln N_a$ by $\ln \bar{s}_a$. Therefore, the condition is:
\be
(\ln \bar{s}_a) \frac{N_a}{\bar{\bar{s}}_a} \ll 1 \quad \text{or} \quad
\frac{N_a}{\bar{\bar{s}}_a} \ll \frac{1}{\ln \bar{s}_a}\,,
\ee
i.e.,
\be\label{schlussbed}
\sum_{\nu=1}^{N_a} \frac{1}{s_{\nu,a}} \ll \frac{1}{\ln \bar{s}_a}\,.
\ee
This means that the $s_{\nu,a}$ must be quite large when compared to their number $N_a$ (i.e., the phase cells must be large compared to their number on the energy surface), and not merely, as assumed in Section~\ref{sec:II4}, large compared to unity. We will investigate later what exactly this means for the distribution of the $s_{\nu,a}$.

We emphasize again the provisional character of our estimates. It is possible that the above stronger assumption on the size of the phase cells is indeed necessary for the ergodic theorem and the $H$-theorem to hold. But maybe it merely arose from the imperfection of our methods of estimation, and in fact the condition $\bar{s}_a \ll 1$ of Section~\ref{sec:II4} is sufficient. It would be of interest to clarify this.

\section{Discussion of the Results}\label{sec:III}

\subsection{}\label{sec:III1}

We sum up the results so far. We have shown:

Let $\psi$ be an arbitrary state, $\psi_t$ the state arising from $\psi$ after time $t$ ($\gtreqqless 0$), $\UUU_\psi$ its micro-canonical ensemble (see Section~\ref{sec:I3}), $\HHH$ the energy operator, $W_{\rho,a}$ its eigenvalues ($a=1,2,\ldots$; $\rho=1,\ldots,S_a$; only those with distinct $a$'s can be distinguished macroscopically, see Section~\ref{sec:I2})---both $\psi$ and $\HHH$ are the exact (rather than the macroscopic) expressions. We assume of $\HHH$ that (for fixed $a$) all $W_{\rho,a}$ are pairwise distinct, and so are all $W_{\rho,a}-W_{\sigma,a}$, $\rho\neq \sigma$, i.e., that $\HHH$ has, within a macroscopically inseparable group of terms, no degeneracies and no resonances with an (imaginary) second equal system.\footnote{Namely, when $W_{\rho,a}-W_{\sigma,s}=W_{\rho',a}-W_{\sigma',a}$ then [the product of] the state $\varphi_{\rho,a}$ in the first system and the state $\varphi_{\sigma',a}$ in the second system has the same total energy as [that of] $\varphi_{\rho',a}$ in the first and $\varphi_{\sigma,a}$ in the second.} (Infrequent violations of these prohibitions can be tolerated.) Then we obtain, in the time average, for the expectation value of any macroscopic observable $\AAA$ and for the entropy:
\begin{align}
M_t \bigl\{ \bigl( E_{\AAA}(\UUU_{\psi}) - E_{\AAA}(\psi_t) \bigr)^2 \bigr\}
&\leq \bar{\eta}^2 \max_{a=1,2,\ldots} \biggl( \sum_{\nu=1}^{N_a} \frac{S_a}{s_{\nu,a}}(\Mb_{\nu,a}+\Nb_{\nu,a})\biggr)\,,\\
M_t \bigl\{ \bigl| S(\UUU_\psi)-S(\psi_t)\bigr|\bigr\}
&\leq \:\quad \max_{a=1,2,\ldots} \biggl( \sum_{\nu=1}^{N_a} \frac{S_a}{s_{\nu,a}}(\Mb_{\nu,a}+\Nb_{\nu,a})\biggr)\,.
\end{align}
(Cf.\ Section~\ref{sec:II3}; it suffices to take the maximum over those $a$ whose (macroscopic) energy surfaces occur in the micro-canonical ensemble $\UUU_\psi$ (i.e., $u_a=(\DDelta_a\psi,\psi)\neq 0$)---in practice this is usually just one $a$. $\bar{\eta}^2$ is the micro-canonical average of $\AAA^2$ and thus a measure of the order of magnitude of the latter.)

The ergodic theorem and the $H$-theorem hold without exception (i.e., for all $\psi$) if
\be\label{cond}
\sum_{\nu=1}^{N_a} \frac{S_a}{s_{\nu,a}} (\Mb_{\nu,a} +\Nb_{\nu,a}) \text{ are small.}
\ee
About the validity of this condition, which involves, apart from $S_a$, $N_a$, $s_{\nu,a}$ (and $\DDelta_a$), also the $\omega_{\lambda,\nu,a}$ (in the $\Mb_{\nu,a}$, $\Nb_{\nu,a}$), we can say this: If
\be\label{cond2}
\sum_{\nu=1}^{N_a} \frac{1}{s_{\nu,a}} \ll \frac{1}{\ln \bar{s}_a}\qquad
\biggl(  \bar{s}_a = \frac{1}{N_a} \sum_{\nu=1}^{N_a} s_{\nu,a}=\frac{S_a}{N_a}\biggr),
\ee
i.e., if the phase cells $\EEE_{\nu,a}$ are large compared to their number on an energy surface $\DDelta_a$, then \eqref{cond} is satisfied for the overwhelming majority of the $\omega_{\lambda,\nu,a}$---i.e., the average over $\omega_{\lambda,\nu,a}$ of $\sum_{\nu=1}^{N_a} (S_a/s_{\nu,a})(\Mb_{\nu,a}+\Nb_{\nu,a})$ is small.\footnote{Note: what we have shown is not that for every given $\psi$ or $\AAA$ the ergodic theorem and the $H$-theorem hold for most $\omega_{\lambda,\nu,a}$ but that for most $\omega_{\lambda,\nu,a}$ they are universally valid, i.e., for all $\psi$ and $\AAA$. The latter is, of course, much more [i.e., much stronger] than the former.}

The real condition \eqref{cond} for the validity of the two theorems can be violated also when \eqref{cond2} holds, i.e., also in this case the macroscopic technique of measurement (the $\omega_{\lambda,\nu,a}$) can be chosen in such a way that the two theorems do not hold. However, for the overwhelming majority of the macroscopic setting, both theorems hold without exceptions (i.e., for all $\psi$ and $\AAA$).

\subsection{}\label{sec:III2}

Let us study \eqref{cond2} more carefully. If all $s_{\nu,a}$ (for a fixed $a$) were of roughly equal size then \eqref{cond2} would amount to $N_a/\bar{s}_a\ll 1/\ln \bar{s}_a$ or $\bar{s}_a/\ln \bar{s}_a \gg N_a$---that is, just a little more than the condition $\bar{s}_a \gg N_a$, which is the statement that the phase cells are large compared to their number on the energy surface. If, on the other hand, the sizes of the $s_{\nu,a}$ are substantially different then we need to be very cautious: already a single $s_{\nu,a}$ that is not $\gg 1$ will have the effect that $\sum_n (1/s_{\nu,a})$ is not $\ll 1$, and thus that our condition \eqref{cond2} is violated. On the other hand, the $s_{\nu,a}$ are very different from one another, as $\ln s_{\nu,a}$ is to be understood as the entropy of the mixture $(1/s_{\nu,a})\EEE_{\nu,a}$ characterizing a general system in the phase cell $\EEE_{\nu,a}$\footnote{This follows from our considerations above or, alternatively, from Boltzmann's definition of entropy, as the phase cell $\EEE_{\nu,a}$ contains $s_{\nu,a}$ states.}---and it suffices to recall the situation in the theory of gases to appreciate that one energy surface will usually contain phase cells with very different entropies. (This fact makes the $H$-theorem a relevant statement.) If the greatest difference in (macroscopically perceptible) entropy among the cells is $\sigma$, so that always
\be
\bigl| \ln s_{\nu,a} - \ln s_{\mu,a} \bigr| \leq \sigma\,,
\ee
then 
\be
s_{\nu,a} \geq \bar{s}_a e^{-\sigma}
\ee
and
\be
\sum_{\nu=1}^{N_a} \frac{1}{s_{\nu,a}} \leq \frac{e^{\sigma}N_a}{\bar{s}_a}\,,
\ee
which leads us to the condition
\be\label{cond3}
\frac{\bar{s}_a}{\ln \bar{s}_a} \gg e^{\sigma} N_a\,.
\ee

This relation shows that no danger arises: since the smallness of $\hbar$ affects the left hand side (because $\bar{s}_a\to \infty$ as $\hbar\to 0$, see Section~\ref{sec:6}) but not the right, \eqref{cond3} will normally be satisfied. We believe that further discussion is not necessary.

\subsection{}
\label{sec:III3}

It remains to discuss the significance of the conditions \eqref{nondeg} and \eqref{nores} on the eigenvalues of $\HHH$ by exemplifying them using the known classical examples and counterexamples to the ergodic theorem and the $H$-theorem.

Let $K$ be a box in which $N$ corpuscles $k_1,\ldots,k_N$ move around, i.e., a gas; we make one of the following two assumptions: either

\begin{itemize}
\item[$\alpha$)] that there is no interaction between the particles, not even collisions (i.e., that they pass through each other); or
\item[$\beta$)] that there are interaction and collisions.
\end{itemize}

In case $\alpha$, it is known that the two theorems do not hold (as any distribution of speeds, not just the Maxwellian, persists for an arbitrarily long time); in case $\beta$, in contrast, one expects the theorems to hold. (The situation is completely analogous for radiation in a cavity with reflecting walls.) How can this behavior be understood from the perspective of our conditions?

Since the $S_a$, $N_a$, $s_{\nu,a}$, and $\EEE_{\nu,a}$ are hardly affected by the difference between $\alpha$ and $\beta$, the condition on $\HHH$ must be relevant. Let us first consider each particle on its own in $K$, and let its energy eigenvalues be $\varepsilon_1,\varepsilon_2,\ldots$.\footnote{We assume that [the particles] $k_1,\ldots,k_N$ are identical and in principle indistinguishable. If they are distinguishable then every [particle] $k_n$ ($n=1,\ldots,N$) possesses a different term spectrum $\varepsilon_{n1},\varepsilon_{n2},\ldots$. The situation is similar to the one we are describing, except that the danger of degeneracy vanishes; $\alpha$ still conflicts with the second condition \eqref{nores} on the eigenvalues of $\HHH$ while $\beta$ does not.} Then, the energy eigenvalues of the total system in $K$ are, in case $\alpha$, the expressions of the form
\be
\sum_{\nu=1}^\infty  z_\nu \varepsilon_\nu
\ee
with $z_\nu=0,1,\ldots$ and $\sum_{\nu=1}^\infty z_\nu =N$, while in case $\beta$ they are slightly modified---the less so the weaker the interaction is. The identity of the particles would lead in general to an $N!$-fold ``permutation degeneracy,''\footnote{In the case $\alpha$. In the case $\beta$, the degrees of degeneracy are the degrees of the irreducible representations of the symmetric group of $N$ elements. Cf.\ \cite{Wig27a,Wig27b,Wig27c}.} and thus to a violation of the first condition \eqref{nondeg} on the energy eigenvalues, but since either Fermi--Dirac or Bose--Einstein statistics apply, i.e., since only wave functions that are anti-symmetric respectively symmetric are admissible \cite{Hei27a,Dir26}, these degeneracies disappear.\footnote{In the case of Fermi--Dirac statistics, only $z_\nu=0,1$ are admissible, but this does not affect our considerations.} Thus, no such difficulty arises. However, in the case $\alpha$ numerous relations of the type excluded by the second condition \eqref{nores} hold:
\be
(\varepsilon_1+\varepsilon_3+\ldots) - (\varepsilon_2+\varepsilon_3+\ldots)=
(\varepsilon_1+\varepsilon_4+\ldots) - (\varepsilon_2+\varepsilon_4+\ldots)
\quad \text{etc.}
\ee
In the case $\beta$ this does not happen because the four above terms of $K$ will be perturbed in very different ways, and, obviously, the absolute magnitude of the perturbation (i.e., of the interaction) does not matter.

Thus, it is the behavior with respect to the condition \eqref{nores} that constitutes the reason for the different character of $\alpha$ and $\beta$.

\appendix

\section{Appendix}
\subsection{}\label{sec:A1}

The properties used in Sections~\ref{sec:II4} and \ref{sec:II5} of the distributions of
\be
\bigl| (\EEE_{\nu,a}\varphi_{\rho,a},\varphi_{\sigma,a}) \bigr|^2 \quad
(\rho\neq\sigma)\quad \text{and}\quad
(\EEE_{\nu,a}\varphi_{\rho,a},\varphi_{\rho,a})
\ee
need to be established. But first we need to explain the sense in which we speak of a statistical distribution.

As we have pointed out in Section~\ref{sec:II4}, everything that depends on $\EEE_{\nu,a}$ ultimately depends on the $\omega_{\lambda,\nu,a}$, and the average we have in mind is the average over these $\omega_{\lambda,\nu,a}$. Since $S_a$, $N_a$, $s_{\nu,a}$ and $\DDelta_a$ are given, they are bound to the condition
\be
\sum_{\nu=1}^{N_a} \sum_{\lambda=1}^{s_{\nu,a}} \PPP_{\omega_{\lambda,\nu,a}}=\DDelta_a
\ee
and determine, in turn, the $\EEE_{\nu,a}$ according to
\be
\sum_{\lambda=1}^{s_{\nu,a}} \PPP_{\omega_{\lambda,\nu,a}} = \EEE_{\nu,a}\,.
\ee
We have also mentioned that all such [orthonormal] systems can be obtained from one of them, say $\overline{\omega}_{\lambda,\nu,a}$ by unitary-linear transformations. Thus, if we choose $\overline{\omega}_{\lambda,\nu,a}$ in whichever way, we can equivalently say that we average over the set of the unitary matrices in $\sum_{\nu=1}^{N_a}s_{\nu,a}=S_a$ dimensions; they map the $\overline{\omega}_{\lambda,\nu,a}$ to the $\omega_{\lambda,\nu,a}$ ($a$ is fixed!). We should denote these matrices by $\{\xi_{\lambda,\nu|\lambda',\nu'}\}$, using for their rows a double index $\lambda,\nu$ and likewise $\lambda',\nu'$ for their columns, corresponding to the notation $\omega_{\lambda,\nu,a}$ and $\overline{\omega}_{\lambda,\nu,a}$ and the relation
\be
\omega_{\lambda,\nu,a} = \sum_{\lambda'=1}^{N_a}\sum_{\nu'=1}^{s_{\nu,a}}
\xi_{\lambda,\nu|\lambda',\nu'} \,\overline{\omega}_{\lambda',\nu',a}\,.
\ee
We prefer, however, to introduce for them the notation $\xi_{\rho|\rho'}$ ($\rho,\rho'=1,\ldots,S_a$). Now we need to explain how to average over the set of the $S_a$-dimensional unitary matrices $\{\xi_{\rho|\rho'}\}$.

We wish to average in a way that does not prefer any reference frame $\overline{\omega}_{\lambda,\nu,a}$ to the others. If $\overline{\overline{\omega}}_{\lambda,\nu,a}$ is another such reference frame and
\be
\overline{\omega}_{\lambda,\nu,a} = \sum_{\lambda'=1}^{N_a} \sum_{\nu'=1}^{s_{\nu,a}} \tilde{\xi}_{\lambda,\nu|\lambda',\nu'} \overline{\overline{\omega}}_{\lambda',\nu',a}\,,
\ee
(we also rewrite $\tilde{\xi}_{\lambda,\nu|\lambda',\nu'}$ as $\tilde{\xi}_{\rho|\rho'}$) then the matrices $\{\xi_{\rho|\rho'}\}$ and $\{\xi'_{\rho|\rho'}\}$ that represent the [orthonormal] system $\omega_{\lambda,\nu,a}$ relative to $\overline{\omega}_{\lambda,\nu,a}$ respectively $\overline{\overline{\omega}}_{\lambda,\nu,a}$ are related according to $\{\xi'_{\rho|\rho'}\} = \{\xi_{\rho|\rho'} \} \{\tilde{\xi}_{\rho|\rho'}\}$ [i.e., $\xi'=\xi\tilde{\xi}$], i.e.,
\be
\xi'_{\rho|\rho''} = \sum_{\rho'=1}^{S_a} \xi_{\rho|\rho'} \tilde{\xi}_{\rho'|\rho''}\,.
\ee
Thus, the procedure of averaging must be invariant under transformations of the above form $\{\xi_{\rho|\rho'}\}\to\{\xi'_{\rho|\rho'}\}$ (for every fixed unitary matrix $\{\tilde{\xi}_{\rho|\rho'}\}$) [i.e., under right multiplication]. Such a procedure of averaging over the unitary group does exist, is uniquely determined by the above requirement, [amounts to integration relative to a measure now known as the Haar measure on the unitary group] and has been specified by Weyl \cite{Wey25}. His general formulas we will not need, as we can reach our goals just by means of the invariance properties of this averaging procedure. We mention that (as shown in \cite{Wey25}) this averaging procedure is also invariant under [left multiplication, i.e.,] the transformation $\{\xi_{\rho|\rho'}\}\to\{\xi''_{\rho|\rho'}\}$ defined by the relation $\{\xi''_{\rho|\rho'}\}=\{\tilde{\xi}_{\rho|\rho'}\}\{\xi_{\rho|\rho'}\}$ [i.e., $\xi''=\tilde{\xi}\xi$], i.e.,
\be
\xi''_{\rho|\rho''} =\sum_{\rho'=1}^{S_a} \tilde{\xi}_{\rho|\rho'} \xi_{\rho'|\rho''}\,.
\ee

Second, for our calculations we simplify the notation. Since the order of the $\nu=1,\ldots,N_a$ is without significance, it suffices to consider $\EEE_{1,a}$. When replacing the two indices $\lambda,\nu$ by one index $\rho$ we can arrange that $(\lambda,1)$ corresponds to $\rho=1,\ldots,s_{1,a}$. Furthermore, we select the reference frame $\overline{\omega}_{\lambda,\nu,a}$: let it be the system of the $\varphi_{\rho,a}$ (where we have also replaced the indices). We thus have that
\be
(\EEE_{1,a}\varphi_{\rho,a},\varphi_{\sigma,a}) 
= \sum_{\tau=1}^{s_{1,a}} (\PPP_{\omega_{\tau,a}} \varphi_{\rho,a},\varphi_{\sigma,a})
\ee 
\be
=\sum_{\tau=1}^{s_{1,a}} (\varphi_{\rho,a},\omega_{\tau,a})(\omega_{\tau,a},\varphi_{\sigma,a}) 
= \sum_{\tau=1}^{s_{1,a}} \xi^*_{\tau,\rho} \xi_{\tau,\sigma}\,.
\ee
Finally, we omit the unnecessary indices $\nu,a$, so that $S_a,N_a,s_{1a},\DDelta_a,\EEE_{1,a},\varphi_{\rho,a},\Mb_{1,a},\Nb_{1,a}$ will be written as $S,N,s,\DDelta,\EEE,\varphi_{\rho},\Mb,\Nb$.\footnote{Note of the translator: Note the difference between $N$ and $\Nb$: $N=N_a$ is the number of macro-states, $\Nb=\Nb_{\nu,a}$ is one of the error bounds.}

Our task is now: As $\{\xi_{\rho,\rho'}\}$ runs through all $S$-dimensional unitary matrices, investigate the distributions, with respect to the [measure corresponding to the] averaging procedure sketched above, of
\be
\bigl| (\EEE\varphi_{\rho},\varphi_{\sigma}) \bigr|^2 
= \Bigl| \sum_{\tau=1}^s \xi^*_{\tau,\rho} \xi_{\tau,\sigma} \Bigr|^2
\quad (\rho\neq\sigma)
\ee
and
\be
(\EEE\varphi_{\rho},\varphi_{\rho}) = \sum_{\tau=1}^s |\xi_{\tau,\rho}|^2\,.
\ee

\subsection{}\label{sec:A2}

We begin with an auxiliary reasoning. We determine the distribution of the values of
\be
\sum_{\rho=1}^s x_\rho^2\,,
\ee
as the vector $\{x_1,\ldots,x_S\}$ runs through the unit sphere
\be\label{spheredef}
\sum_{\rho=1}^S x_\rho^2=1\,,
\ee
at first with real $x_{\rho}$. That is, we determine $W(u)$, where $W(u)\,du$ is the (geometric) probability for
\be\label{Wdef}
u \leq \sum_{\rho=1}^s x_\rho^2 \leq u+du
\ee
($0\leq u\leq 1$).\footnote{This amounts to determining the surface area of the $s$-dimensional calotte on the $S$-dimensional unit sphere.} Simple geometrical considerations that we need not reproduce here show that $W(u)$ is proportional to
\be
u^{s/2-1}(1-u)^{(S-s)/2-1}\,,
\ee
where the proportionality factor needs to be determined from
\be
\int_0^1 W(u)\, du=1\,.
\ee
Now, if we allow $x_1,\ldots,x_S$ to be complex and consider
\be
u\leq \sum_{\rho=1}^s |x_{\rho}|^2 \leq u+du
\ee
instead of \eqref{Wdef} and
\be
\sum_{\rho=1}^S |x_\rho|^2=1
\ee
instead of \eqref{spheredef}, then we realize that the problem has not changed as we can regard the real and imaginary parts of the $x_\rho$ as real Cartesian coordinates. Thus, we only need to replace $s,S$ by $2s,2S$, so $W(u)$ becomes proportional to
\be\label{W}
u^{s-1}(1-u)^{S-s-1}\,,
\ee
and the proportionality factor can be determined from the normalization condition to be
\be
\frac{(S-1)!}{(s-1)!(S-s-1)!}\,.
\ee 
Therefore, 
\begin{align}
&\text{the average of }\Bigl(\sum_{\rho=1}^s |x_\rho|^2 \Bigr)^n\nonumber\\
&= \int_0^1 \frac{(S-1)!}{(s-1)!(S-s-1)!} u^{s-1}(1-u)^{S-s-1} u^n du\\
&= \frac{(S-1)!}{(s-1)!(S-s-1)!} \int_0^1 u^{s+n-1}(1-u)^{S-s-1} du\\
&= \frac{(S-1)!}{(s-1)!(S-s-1)!} \frac{(s+n+1)!(S-s-1)!}{(S+n-1)!}\\
&= \frac{s(s+1)\cdots(s+n-1)}{S(S+1)\cdots(S+n-1)}\,.
\end{align}

\subsection{}\label{sec:A3}

We return to the unitary matrix $\xi_{\rho|\rho'}$ and introduce the abbreviation 
\be
e_{\rho,\sigma} = \sum_{\tau=1}^s \xi^*_{\tau,\rho}\xi_{\tau,\sigma}\,.
\ee
For the reasons described in Appendix~\ref{sec:A1}, all $e_{\rho,\sigma}$ ($\rho\neq\sigma$) have the same probability distribution, and likewise all $e_{\rho,\rho}$.\footnote{The interchange of columns and that of rows belongs to the transformations there [under which the Haar measure is invariant].}

In
\be
e_{\rho,\rho} = \sum_{\tau=1}^s |\xi_{\tau,\rho}|^2\,,
\ee
only the $\rho$-th column of $\{\xi_{\rho|\rho'}\}$ appears, over which can be averaged in the same way as we averaged over the unit sphere in Appendix~\ref{sec:A2} [i.e., whose distribution is uniform on the unit sphere] (this follows easily from the invariance properties of the averaging procedure). Thus (denoting the average by $\M$),
\be
\M(e_{\rho\rho}) = \frac{s}{S}\,,\quad
\M(e_{\rho\rho}^2) = \frac{s(s+1)}{S(S+1)}\,,
\ee
\begin{align}
\M\biggl( \Bigl(e_{\rho\rho}-\frac{s}{S} \Bigr)^2 \biggr)
&= \M(e_{\rho\rho}^2) - \frac{2s}{S} \M(e_{\rho\rho}) + \frac{s^2}{S^2}\\
&= \frac{s(s+1)}{S(S+1)}-\frac{s^2}{S^2} = \frac{s(S-s)}{S^2(S+1)}\,.
\end{align}
Furthermore, $\EEE^2=\EEE$ implies
\be
e_{\rho\rho} = \sum_{\sigma=1}^S |e_{\rho\sigma}|^2
=e_{\rho\rho}^2 + \sum_{\substack{\sigma=1\\\sigma\neq \rho}}^S |e_{\rho\sigma}|^2\,.
\ee
Due to the equality of the $\M(|e_{\rho\sigma}|^2)$ ($\rho\neq\sigma$), we have that
\begin{align}
\M(|e_{\rho\sigma}|^2) 
&= \frac{1}{S-1} \bigl(\M(e_{\rho\rho}) - \M(e_{\rho\rho}^2) \bigr)\\
&= \frac{1}{S-1} \Bigl( \frac{s}{S} - \frac{s(s+1)}{S(S+1)} \Bigr) 
= \frac{s(S-s)}{S(S^2-1)}\,.
\end{align}
The averages used in Section~\ref{sec:II4} have thus been determined in agreement with the values used there.

Now we turn to investigating the distributions of
\be
|e_{\rho\sigma}|^2 \:\:(\rho\neq\sigma)\quad\text{and}\quad
\Bigl(e_{\rho\rho}-\frac{s}{S} \Bigr)^2
\ee
in order to determine the averages of $\Mb$ and $\Nb$ as in Section~\ref{sec:II5}.

\subsection{}\label{sec:A4}

The latter problem is the easier one. We know already that $u\leq e_{\rho\rho}\leq u+du$ (with $0\leq u \leq 1$) has probability $W(u)\, du$ (see Appendix~\ref{sec:A2}). Let $a$ be a positive number with $a\ll s^2/S^2$; then the probability of 
\be
(e_{\rho\rho}-s/S)^2 \geq a
\ee
(note that the left hand side is certainly less than or equal to 1, as $0\leq e_{\rho\rho}\leq 1$) is
\[
\biggl(\int\limits_0^{s/S-\sqrt{a}} + \int\limits_{s/S+\sqrt{a}}^1 \biggr) W(u)\, du
\]
\be\label{expr3}
=\frac{(S-1)!}{(s-1)!(S-s-1)!} \biggl( \int\limits_0^{s/S-\sqrt{a}} + \int\limits_{s/S+\sqrt{a}}^1\biggr) u^{s-1} (1-u)^{S-s-1}du\,.
\ee
The derivative of the logarithm of the integrand equals
\be
\frac{s-1}{u}-\frac{S-s-1}{1-u} = \frac{1}{u(1-u)} \bigl( [s-1]-[S-2]u\bigr)\,,
\ee
so the integrand increases when $u$ approaches $(s-1)/(S-2)$ from either side. This point lies to the left of $s/S$, in fact by an amount of 
\footnote{Note of the translator: In the German original, Eq.~\eqref{difference} is misprinted as
\[
\frac{s}{S}-\frac{s-1}{S-1} + \frac{S-2s}{S(S-1)} \leq \frac{1}{S}\,.
\]
}
\be\label{difference}
\frac{s}{S}-\frac{s-1}{S-2} = \frac{S-2s}{S(S-2)} \leq \frac{1}{S}\,,
\ee
and thus still lies in the interval $s/S \pm \sqrt{a}$ provided $a\geq 1/S^2$. Therefore, within the domain of integration, the integrand assumes its maximum at $u=s/S\pm\sqrt{a}$ (we will not try to find out at which of the two values). We can thus estimate the entire expression \eqref{expr3} as being
\be\label{expr4}
\leq \frac{(S-1)!}{(s-1)!(S-s-1)!} \Bigl(\frac{s}{S}\pm \sqrt{a}\Bigr)^{s-1} \Bigl( 1-\frac{s}{S} \mp \sqrt{a}\Bigr)^{S-s-1}\,.
\ee
Now we use the assumption $1\ll s\ll S$, which implies that the first factor is, by Stirling's formula, approximately equal to\footnote{Note of the translator: In the German original, the second exponent in this expression is misprinted as $S-s$, and the factor $1/e=\exp(-1)$, which is as irrelevant as the $\sqrt{2\pi}$ to the purpose at hand, is missing here and in the following.}
\be
\frac{1}{e}\sqrt{\frac{s}{2\pi}} \Bigl(\frac{s}{S}\Bigr)^{-s} \Bigl( 1-\frac{s}{S}\Bigr)^{s-S}\,,
\ee
while the second is approximately equal to
\be
\frac{S}{s} \Bigl(\frac{s}{S} \pm \sqrt{a}\Bigr)^s \Bigl( 1-\frac{s}{S} \mp \sqrt{a}\Bigr)^{S-s}\,.
\ee
The entire expression \eqref{expr4} is therefore approximately equal to
\begin{align}
&\frac{S}{e\sqrt{2\pi s}} \Bigl(1\pm \frac{S}{s}\sqrt{a} \Bigr)^s
\Bigl( 1\mp \frac{S}{S-s} \sqrt{a}\Bigr)^{S-s}\\
&= \frac{S}{e\sqrt{2\pi s}} \exp\biggl( s\ln \Bigl(1\pm \frac{S}{s}\sqrt{a}\Bigr) + (S-s) \ln \Bigl( 1\mp \frac{S}{S-s} \sqrt{a} \Bigr) \biggr)\,.
\end{align}
The exponent is [because $\ln(1+x)\leq x-x^2/2+x^3/3$ and $\ln(1+x)\leq x$] less than or equal to
\begin{align}
&\pm s \frac{S\sqrt{a}}s - s\frac{S^2a}{2s^2} \pm s\frac{S^3a\sqrt{a}}{3s^3} \mp (S-s) \frac{S}{S-s} \sqrt{a}\\
&= -\frac{S^2a}{2s} \pm \frac{S^3a\sqrt{a}}{3s^2}\,.
\end{align}
Since $s\sqrt{a}/S \ll 1$, the second term is small compared to the first, and thus the expression \eqref{expr4} is
\be
\lesssim \frac{S}{e\sqrt{2\pi s}} e^{-\Theta\frac{S^2a}{2s}}
\ee
($\Theta$ some number less than 1).

This concerned the probability of $(e_{\rho\rho}-s/S)^2\geq a$ for a fixed $\rho=1,\ldots,S$; the probability that this event occurs for some $\rho$, i.e., the probability of
\be
\Nb=\max_{\rho=1}^S \Bigl(e_{\rho\rho}-\frac{s}{S} \Bigr)^2 \geq a\,,
\ee
is at most $S$ times larger, and thus
\be
\lesssim \frac{S^2}{e\sqrt{2\pi s}} e^{-\Theta\frac{S^2a}{2s}}
\ee
Now we estimate the average of $\Nb$ in two parts: for values in $[0,a]$, the probability is at most 1, for values in $[a,1]$ we have the above bound. Therefore,
\be
\M(\Nb) \lesssim a + \frac{S^2}{e\sqrt{2\pi s}} e^{-\Theta\frac{S^2a}{2s}}\,.
\ee
Here, $a$ can be chosen to be any number such that $a\geq 1/S^2$ and $a\ll s^2/S^2$; we choose
\be
a= \frac{8s\ln S}{\Theta S^2}\,.
\ee
(This satisfies everything, provided $s\gg \ln S$, which must be the case anyway by condition \eqref{schlussbed}.\footnote{From $\sum_{\nu=1}^{N_a} 1/s_{\nu,a} \ll 1/\ln \bar{s}_a$ follows $s_{\nu,a}\gg \ln \bar{s}_a$. Put differently, see \eqref{expr5}, 
$N_a\ln S_a /\bar{\bar{s}}_a\ll 1$, [or, equivalently,] $S_a \ln S_a/(\bar{s}_a\bar{\bar{s}}_a)\ll 1$ so a fortiori $S_a/\bar{s}_a^2 \leq 1$, $\bar{s}_a\geq \sqrt{S_a}$, $\ln s_a \geq \tfrac{1}{2} \ln S_a$. Thus, we have that $s_{\nu,a}\gg \ln S_a$, i.e., $s\gg \ln S$.}) Our upper bound thus becomes
\be
\frac{8s\ln S}{\Theta S^2}+\frac{S^2}{e\sqrt{2\pi s}}e^{-4\ln S}=
\frac{8s\ln S}{\Theta S^2}+\frac{1}{e\sqrt{2\pi s}S^2}\sim 
\frac{8s\ln S}{\Theta S^2}\,.
\ee
Thus, if the premise $1\ll s\ll S$ is satisfied to a sufficient extent, the above average is certainly less than or equal to $9s\ln S/S^2$.

\subsection{}
\label{sec:A5}

It remains to discuss the distribution of $|e_{\rho\sigma}|^2$ ($\rho\neq \sigma$). We denote the $\rho$-th and the $\sigma$-th column of $\{\xi_{\tau|\tau'}\}$ by $\xi=\{\xi_{1|\rho},\ldots,\xi_{S|\rho}\}$ and $\eta=\{\xi_{1|\sigma},\ldots,\xi_{S|\sigma}\}$; in addition, let $\tilde{\xi}=\{\xi_{1|\rho},\ldots,\xi_{s|\rho},0,\ldots,0\}$. For such vectors $\zeta=\{\zeta_1,\ldots,\zeta_S\}$, $\chi=\{\chi_1,\ldots,\chi_S\}$ we will also use the notation
\be
(\zeta,\chi) =\sum_{\tau=1}^S \zeta_\tau \chi^*_\tau\,,\quad
|\zeta| = \sqrt{(\zeta,\zeta)} = \sqrt{\sum_{\tau=1}^S |\zeta_\tau|^2}\,.
\ee
We have that
\be
|e_{\rho\sigma}|^2 = |(\tilde{\xi},\eta)|^2\,,
\ee
where the vectors $\xi,\eta$, being columns of a unitary matrix, are subject to the conditions $|\xi|=1$, $|\eta|=1$, $(\xi,\eta)=0$ (i.e., both lie on the unit sphere and are orthogonal to each other).

We decompose $\tilde{\xi}$ into a component parallel to $\xi$ and one orthogonal to $\xi$:
\be
\tilde{\xi} = (\tilde{\xi},\xi) \xi + \tilde{\tilde{\xi}}\,.
\ee
Then we can just as well write
\be
|e_{\rho\sigma}|^2 = \bigl| (\tilde{\tilde{\xi}},\eta)\bigr|^2\,.
\ee
When keeping $\xi$ (and $\tilde{\xi},\tilde{\tilde{\xi}}$) fixed, we thus have two vectors $\tilde{\tilde{\xi}},\eta$ orthogonal to $\xi$, of which the first is fixed and the second can vary freely on the surface of a $(S-1)$-dimensional unit ball. We introduce an arbitrary $(S-1)$-dimensional Cartesian coordinate system for this [subspace], let
\be
\eta=(y_1,\ldots,y_{S-1})\,.
\ee
From the unitary invariance of our averaging procedure follows that the procedure amounts (for fixed $\xi=\{\xi_{1|\rho}, \ldots, \xi_{S|\rho}\}$) exactly to averaging $\eta$ over the $(S-2)$-dimensional unit sphere\footnote{Note of the translator: The German original literally says here: over the $(S-1)$-dimensional unit ball.} as described in Appendix~\ref{sec:A2}. Moreover, due to the unitary invariance, the only thing that matters about $\tilde{\tilde{\xi}}$ is its length $|\tilde{\tilde{\xi}}|$, so we can replace it by
\be
\tilde{\tilde{\tilde{\xi}}}=\bigl\{|\tilde{\tilde{\xi}}|,0,\ldots,0\bigr\}
\ee
(in $S-1$ dimensions). That is why we first aim at determining the distribution of
\be\label{expr6}
\bigl| (\tilde{\tilde{\tilde{\xi}}},\eta) \bigr|^2 =|\tilde{\tilde{\xi}}|^2 \, |y_1|^2
\ee
for [random $\eta$ with] 
\be
|\eta|^2 = \sum_{\pi=1}^{S-1} |y_\pi|^2=1\,.
\ee 
That \eqref{expr6} lies in $[u,u+du]$ ($0\leq u \leq |\tilde{\tilde{\xi}}|^2$) means that 
\be
\frac{u}{|\tilde{\tilde{\xi}}|^2}\leq |y_1|^2 \leq \frac{u}{|\tilde{\tilde{\xi}}|^2} + \frac{du}{|\tilde{\tilde{\xi}}|^2}\,,
\ee
which has probability
\be
W\Bigl( \frac{u}{|\tilde{\tilde{\xi}}|^2}\Bigr) \frac{du}{|\tilde{\tilde{\xi}}|^2}\,,
\ee
where $W$ is given by \eqref{W} with $s,S$ replaced by $1,S-1$. Thus, the coefficient of $du$ is:\footnote{Note of the translator: In the German original, the formula corresponding to \eqref{expr7} has $S-1$ instead of $S-2$ and $S-2$ instead of $S-3$. This mistake propagates through all further formulas in the German original but does not affect the final result. Here and in the following, we give the correct exponents.}
\be\label{expr7}
\frac{S-2}{|\tilde{\tilde{\xi}}|^{2(S-2)}}
\bigl( |\tilde{\tilde{\xi}}|^2 - u \bigr)^{S-3}
\ee

While we had kept $\xi$ fixed up to now, we will now average \eqref{expr7} (of course, in the sense of Appendix~\ref{sec:A2} [i.e., using to a uniform distribution of $\xi$]) over the ($S$-dimensional) unit sphere. The expression for the distribution of $|e_{\rho\sigma}|^2$ for given $\xi$ depends only on $|\tilde{\tilde{\xi}}|^2$, and (since $\tilde{\xi}$ is orthogonal to both $\xi-\tilde{\xi}$ and $\tilde{\tilde{\xi}}=\tilde{\xi}-(\xi,\tilde{\xi})\tilde{\xi}$) we have that
\begin{align}
|\tilde{\xi}|^2 &= (\tilde{\xi},\tilde{\xi}) = (\xi,\tilde{\xi})\,,\\
|\tilde{\xi}|^2 &= \bigl| (\xi,\tilde{\xi})\xi\bigr|^2 + |\tilde{\tilde{\xi}}|^2 = |\tilde{\xi}|^4 + |\tilde{\tilde{\xi}}|^2\,,\\
&|\tilde{\tilde{\xi}}|^2 = |\tilde{\xi}|^2 (1-|\tilde{\xi}|^2)\,.
\end{align}
Since $\xi=\{\xi_{1|\rho},\ldots,\xi_{S|\rho}\}$ varies on the unit sphere, the event
\be
w\leq |\tilde{\xi}|^2 \leq w+dw
\ee
($0\leq w \leq 1$), i.e.,
\be
w\leq \sum_{\tau=1}^s |\xi_{\tau|\rho}|^2 \leq w+dw\,,
\ee
has probability
\be
\frac{(S-1)!}{(s-1)!(S-s-1)!} w^{s-1} (1-w)^{S-s-1} \, dw\,.
\ee
In order to obtain the total probability density of $|e_{\rho\sigma}|^2$ at $u$, we thus need to integrate
\begin{align}
&\frac{(S-1)!}{(s-1)!(S-s-1)!} w^{s-1} (1-w)^{S-s-1} \: \times\nonumber\\
&\times \:
\frac{S-2}{\bigl(w(1-w)\bigr)^{S-2}}
\bigl( w(1-w) - u \bigr)^{S-3} \, dw \nonumber\\
&= \frac{(S-1)!(S-2)}{(s-1)!(S-s-1)!} 
\frac{ \bigl( w(1-w) - u \bigr)^{S-3} }{w^{S-s-1}(1-w)^{s-1}} dw 
\label{expr9}
\end{align}
over those $w\in [0,1]$ with $u\leq w(1-w)$. As a consequence, only values in $[0,\tfrac{1}{4}]$ can arise for $u$. We now determine the probability of $|e_{\rho\sigma}|^2\geq a$ (with $0\leq a \leq \tfrac{1}{4}$), and to this end we need to integrate \eqref{expr9} over those $u,w$ with $a\leq u \leq w(1-w)$, i.e., over those $u,w$ with
\be
\tfrac{1}{2} - \sqrt{\tfrac{1}{4}-a}\leq w \leq \tfrac{1}{2} + \sqrt{\tfrac{1}{4}-a}\,, \quad
a\leq u \leq w(1-w)\,.
\ee
We can carry out the integration over $u$:\footnote{Note of the translator: In the German original, \eqref{expr8} contains an inconsistency (the numerators of the integrands in the left and right hand sides have equal exponents) that partly compensates the mistake about exponents in \eqref{expr7}.}
\begin{align}
&\quad \frac{(S-1)!(S-2)}{(s-1)!(S-s-1)!} 
\:\:\int\limits_{\frac12 -\sqrt{\frac14 -a}}^{\frac12 +\sqrt{\frac14 -a}}\:\:
\int\limits_a^{w(1-w)}
\frac{ \bigl( w(1-w) - u \bigr)^{S-3} }{w^{S-s-1}(1-w)^{s-1}} du\, dw\nonumber\\ 
&=\frac{(S-1)!}{(s-1)!(S-s-1)!} 
\:\:\int\limits_{\frac12 -\sqrt{\frac14 -a}}^{\frac12 +\sqrt{\frac14 -a}}\:\:
\frac{ \bigl( w(1-w) - a \bigr)^{S-2} }{w^{S-s-1}(1-w)^{s-1}} dw\,. \label{expr8}
\end{align}
We decompose the integral into two parts,
\[
\int\limits_{\frac12}^{\frac12 -\sqrt{\frac14 -a}} \quad \text{and}\quad
\int\limits_{\frac12 +\sqrt{\frac14 -a}}^{\frac12}\,,
\]
and introduce the new variable $x$ according to
\be
\tfrac12+\sqrt{\tfrac14 -x}=w\,,\quad \text{respectively}\quad
\tfrac12-\sqrt{\tfrac14-x}=w\,.
\ee
In both cases we have that $x=w(1-w)$, and in both cases $x$ runs from $a$ to $\tfrac14$. Combining both integrals, we arrive at
\begin{multline}
\frac{(S-1)!}{(s-1)!(S-s-1)!} \int\limits_a^{\frac14} (x - a)^{S-2} 
\Bigl[ \bigl(\tfrac12 + \sqrt{\tfrac14-x}\bigr)^{-(S-s-1)}
\bigl(\tfrac12-\sqrt{\tfrac14 -x} \bigr)^{-(s-1)}+\\
\bigl(\tfrac12-\sqrt{\tfrac14-x}\bigr)^{-(S-s-1)} 
\bigl(\tfrac12+\sqrt{\tfrac14-x}\bigr)^{-(s-1)}\Bigr]
\frac{dx}{2\sqrt{\tfrac14-x}}\,.
\end{multline}

Finally, we introduce the new variable
\be
y=\frac{x-a}{\tfrac14-a}\,,
\ee
which runs from 0 to 1. The above expression then becomes\footnote{Note of the translator: In the German original, a factor $(1-4a)^{-1/2}$ is missing here and in the following equations. This mistake does not affect the final result.}
\begin{multline}
\frac{(1-4a)^{S-2-\frac12}(S-1)!}{2^{S-2}(s-1)!(S-s-1)!}  \int_0^1 y^{S-2} \:\times \\
\times\:\Bigl[ (1 + \sqrt{1-4a}\sqrt{1-y})^{-(S-s-1)}(1-\sqrt{1-4a}\sqrt{1-y})^{-(s-1)}+\\
(1-\sqrt{1-4a}\sqrt{1-y})^{-(S-s-1)} (1+\sqrt{1-4a}\sqrt{1-y})^{-(s-1)}\Bigr]
\frac{dy}{\sqrt{1-y}}\,.\label{expr10}
\end{multline}
Once we divide this probability by $(1-4a)^{S-2-\frac12}$, only the square bracket depends on $a$. As we will show, the square bracket increases as $a\to 0$, and thus so does the quotient [i.e., \eqref{expr10}/$(1-4a)^{S-2-\frac12}$]. Since for $a=0$, \eqref{expr10} is 1, as well as $(1-4a)^{S-2-\frac12}=1$, this implies that the quotient is always less than or equal to 1, and thus 
\be
\eqref{expr10} \leq (1-4a)^{S-2-\frac12} \leq e^{-4a(S-2-\frac12)}\,.
\ee

As $a\to 0$, $\sqrt{1-4a}\sqrt{1-y}$ tends, monotonically increasingly, to $\sqrt{1-y}$, so it suffices to show that 
\be
\bigl[ (1 + t)^{-(S-s-1)}(1-t)^{-(s-1)}+(1-t)^{-(S-s-1)} (1+t)^{-(s-1)}\bigr]
\ee
is an increasing function of $t$ if $t>0$ [and $t<1$]. Indeed, its derivative
\begin{multline}\label{expr11}
(1 + t)^{-(S-s-1)}(1-t)^{-(s-1)} \Bigl(\frac{s-1}{1-t}-\frac{S-s-1}{1+t} \Bigr)+\\
(1-t)^{-(S-s-1)} (1+t)^{-(s-1)} \Bigl( \frac{S-s-1}{1-t} -\frac{s-1}{1+t} \Bigr)
\end{multline}
is positive if (we set $\displaystyle z= \frac{1+t}{1-t}>1$) [as we see by multiplying \eqref{expr11} by $(1+t)^{S+1}>0$]
\be
z^{s+1} \bigl((s-1)z-(S-s-1) \bigr)
+z^{S-s-1} \bigl( (S-s-1)z -(s-1) \bigr)>0\,,
\ee
but this expression is obviously greater than
\be
z^{s+1} \bigl((s-1)-(S-s-1) \bigr)
+z^{S-s-1} \bigl( (S-s-1) -(s-1) \bigr)
=(z^{S-s-1}-z^{s+1}) (S-2s)\geq 0
\ee
[because $z>1$ and $S\geq 2s+2$]. Thus, we have verified the above bound for the probability of $|e_{\rho\sigma}|^2\geq a$ for a fixed pair $\rho\neq \sigma$, $\rho,\sigma=1,\ldots,S$. The probability that this occurs for any such $\rho,\sigma$, i.e., the probability of
\be
\Mb=\max^S_{\substack{\rho,\sigma=1\\\rho\neq \sigma}}\bigl( |e_{\rho\sigma}|^2 \bigr) \geq a\,,
\ee
is larger by at most a factor of $S(S-1)/2$ (because of $e_{\rho\sigma}=e_{\sigma\rho}^*$ it suffices to consider $\rho<\sigma$), and thus is less than or equal to
\be
\frac{S(S-1)}{2}e^{-4a(S-2-\frac12)}\,.
\ee
The average of $\Mb$ we estimate again in two parts: for values in $[0,a]$ the probability is certainly $\leq 1$, for values in $[a,\tfrac14]$ we have the above bound. Therefore:
\be
\M(\Mb) \leq a+ \frac{S(S-1)}{8}e^{-4a(S-2-\frac12)}\,.
\ee
For $a$ we can choose any number $\geq 0$, $\ll 1$, we set
\be
a=\frac34 \frac{\ln S}{S}\,.
\ee
(This fulfills all requirements because of $S\gg 1$.) Our upper bound thus becomes
\be
\frac{3}{4} \frac{\ln S}{S} + \frac{S(S-1)}{8}e^{-3\ln S \frac{S-2-\frac12}{S}}
\sim \frac{3}{4} \frac{\ln S}{S} + \frac{S^2}{8}e^{-3\ln S}
= \frac{3}{4} \frac{\ln S}{S} + \frac{1}{8S}
\sim \frac{3}{4} \frac{\ln S}{S} \,.
\ee
Thus, if the premise $S\gg 1$ is satisfied to a sufficient extent then the above average is less than or equal to $\ln S/S$.

This completes the proof of the desired estimates.

\end{document}